\newif\ifINCLUDEFIGS 
\INCLUDEFIGSfalse
\INCLUDEFIGStrue

\newif\ifFIGSATEND
\FIGSATENDfalse

\newif\ifkeypoint 
\keypointtrue
\keypointfalse

\newif\ifseparate
\separatetrue
\newif\ifTTD

\newif\ifFIG
\FIGtrue

\ifTTD

==
\fi

\documentclass[a4paper,11pt]{article} 

\usepackage{amsmath}
\usepackage{graphicx}
\usepackage{setspace}
\doublespace
\usepackage{fullpage}
\usepackage{latexsym}
\usepackage{amsfonts}

\usepackage{epsf}
\usepackage{url}
	\usepackage{subfig} 

\usepackage[dvipsnames,usenames]{color}

\usepackage{ulem}

\usepackage{soul}
\usepackage{amsthm}
\usepackage{thmtools}

\usepackage{tikz}
\tikzstyle{every picture}+=[font=\sffamily] 
\usetikzlibrary{decorations.pathmorphing} 
\usetikzlibrary{shapes.geometric,calc}
\usetikzlibrary{shapes,arrows}
\usepackage{lipsum}

\usepackage{setspace,lipsum}
\newcommand{\keypointwidth}{0.9}

\tikzset{
    >=stealth'
	}
\usetikzlibrary{decorations.pathmorphing} 
\usetikzlibrary{shapes.geometric,calc}
\usetikzlibrary{shapes,arrows}
\tikzstyle{int}=[draw, minimum size=2em]
\tikzstyle{init} = [pin edge={to-,thick,black}]
\usetikzlibrary{arrows,decorations.pathmorphing,backgrounds,positioning,fit,petri}

\newlength{\RoundedBoxWidth}
\newsavebox{\GrayRoundedBox}

\newenvironment{KeyPointBox}[1][\dimexpr\textwidth-4.5ex]%
   {\setlength{\RoundedBoxWidth}{\dimexpr#1}
    \begin{lrbox}{\GrayRoundedBox}
       \begin{minipage}{\RoundedBoxWidth}}%
   {   \end{minipage}
    \end{lrbox}
    \begin{center}
    \begin{tikzpicture}%
       \draw node[draw=black,fill=black!5,rounded corners,%
             inner sep=2ex,text width=\RoundedBoxWidth]%
             {\usebox{\GrayRoundedBox}};
    \end{tikzpicture}
    \end{center}}
    
   {\setlength{\RoundedBoxWidth}{\dimexpr#1}
    \begin{lrbox}{\GrayRoundedBox}
       \begin{minipage}{\RoundedBoxWidth}}%
   {   \end{minipage}
    \end{lrbox}
    \begin{center}
    \begin{tikzpicture}%
       \draw node[draw=black,fill=black!5,rounded corners,%
             inner sep=2ex,text width=\RoundedBoxWidth]%
             {\usebox{\GrayRoundedBox}};
    \end{tikzpicture}
    \end{center}}

\newcommand{\rvs}{x}
\newcommand{\rvr}{y}

\renewcommand{\em}{\it}

\newcommand{\edit}[1]{{#1}}
\newcommand{\editp}[1]{{#1}}
\newcommand{\delete}[1]{\unskip}
\newcommand{\deletep}[1]{\unskip}
\newcommand{\CO}[1]{\unskip}
\newcommand{\query}[1]{\unskip}

\newcommand{\jsubsection}[1]{\vspace{0.05in}\ni  {\bf #1}.}

\newcommand{\eg}{e.g.}

\renewcommand{\ni}{\noindent}

\newcommand{\xt}{x_{t}}
\newcommand{\xh}{x_{h}}

\newcommand{\jlog}	{\log}

\newcommand{\var}{\mathrm{var}}

\newcommand{\lt}{<}

\newcommand{\be}        { \begin{equation}  }
\newcommand{\ee}        { \end{equation}	}
\newcommand{\bea}       { \begin{eqnarray}  }
\newcommand{\eea}       { \end{eqnarray}    }

\newcommand{\nn}	{\nonumber}

\newcommand{\jem}	{\it}

\newcommand{\bq}	{\begin{quote} \jem}
\newcommand{\eq}	{\end{quote}}

\newcommand{\Y}	{r}

\newcommand{\X}	{s}

\renewcommand{\X}{x}
\renewcommand{\rvs}{x}

\newcommand{\nc}{m} 
\newcommand{\s}{x}

\newcommand{\bits}{\:{\rm bits}}

\newcommand{\bitss}{\:{\rm bits/s}}

\newcommand{\nl}{\newline}
\newcommand{\jvsblack}{}

\newcommand{\noise}{\eta}

\newcommand{\JVSdelete}[1]{\unskip}

\renewcommand{\r}{y}

\begin{document}

\thispagestyle{empty}

\title{
{\bf 
Information Theory:   A Tutorial Introduction
}
}

\author{
James V Stone, 
Psychology Department, University of Sheffield, England.\\
j.v.stone@sheffield.ac.uk\\
File: main\_InformationTheory\_JVStone\_v4.tex
}

\date{}
\maketitle

%
%

\onehalfspacing

\begin{abstract}
Shannon's mathematical theory of communication defines fundamental limits on how much information can be transmitted between the different components of any man-made or biological system. This paper is an informal but rigorous introduction to the main ideas implicit in Shannon's theory. An annotated reading list is provided for further reading.  
\end{abstract}

\section{Introduction}

\index{Shannon, C}%

\ifFIG
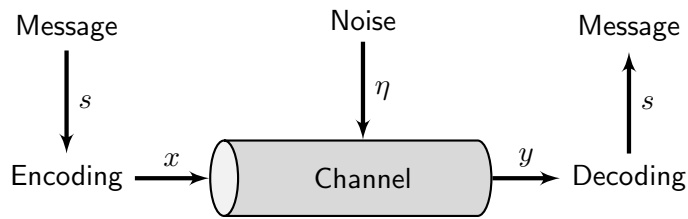
\begin{figure}[b] 
\begin{center}
\begin{tikzpicture}
[node distance = 0.3cm, auto, >=latex']

	\node [ultra thick] (t) [node distance=1.5cm]{Encoding};

	\node (m) [above of=t, node distance = 2cm]{Message};

	\path[->,ultra thick] (m) edge node [right] {$s$}         (t);
	
	 \node [ultra thick,right of =t,cylinder,draw=black,thick,aspect=0.7,minimum height=3.7cm,minimum width=1.0cm,shape border rotate=180,cylinder uses custom fill, cylinder body fill=gray!30,cylinder end fill=gray!10,node distance = 3.9cm] (zz) {Channel};

	\node [ultra thick] (n) [above of=zz, node distance = 2.1cm] {Noise};
	\node [ultra thick] (op) [right of=zz, node distance = 3.5cm] {Decoding};
	
	\path[->,ultra thick] (n) edge node [right] {$\eta$} (zz);
	\path[->, ultra thick] (t) edge node [above] {$x$} (zz);
	\path[->, ultra thick] (zz) edge node [above] {$y$} (op);

	\node (blank) [ultra thick][above of=op, node distance = 2cm]{Message};

	\path[->, ultra thick] (op) edge node [right] {$s$} (blank);
\end{tikzpicture}
\caption{The communication channel.  A message (data) is encoded before being used as input to a  communication channel, 
which adds noise. The channel output is decoded by a receiver to recover the message. 
}
\label{figchannel1}
\end{center}
\end{figure}
\fi

\index{Shannon, C}
In 1948, Claude Shannon published a paper called\delete{,} \delete{``} {\em A Mathematical Theory of Communication}\delete{''}\cite{ShannonPaper}. 
This paper heralded a transformation in our \delete{conception} \edit{understanding} \query{`understanding' might work better for an international audience} of information. Before Shannon's paper, information had been viewed as a kind of poorly defined miasmic fluid. But after Shannon's paper, it became apparent that information is a well\editp{-}defined\delete{,} and\edit{,} above all, {\em measurable} quantity. Indeed, as noted by Shannon, 

\begin{quote}
{\em A basic idea in information theory is that information can be treated very much like a physical quantity, such as mass or energy.}\\
Caude Shannon, 1985. 
\end{quote}

Information theory defines definite, unbreachable limits on precisely how much information can be communicated
between any two components of any system, whether this system is man-made or natural.
\index{theorem} %
The theorems of information theory  are so important that they deserve to be regarded as the {\em laws} of information\cite{ShannonWeaverBook, REZA_INFO_THEORY,StoneInformationBook2014}. 
\index{encoding}%
\index{channel capacity}%
\index{communication channel}%

The basic laws of information  
can be summarised as follows. 
For any communication channel (Figure \ref{figchannel1}):  
1) there is a definite upper limit, the {\em channel capacity}, to the amount of information that can be communicated through that channel, 2) this limit shrinks as the amount of noise in the channel increases,
3) this limit can very nearly be reached by judicious packaging, or {encoding}, of data.

\ifFIG
\begin{figure}[b!] %
\begin{center}
\includegraphics[ width =0.6 \textwidth, angle=0 ] {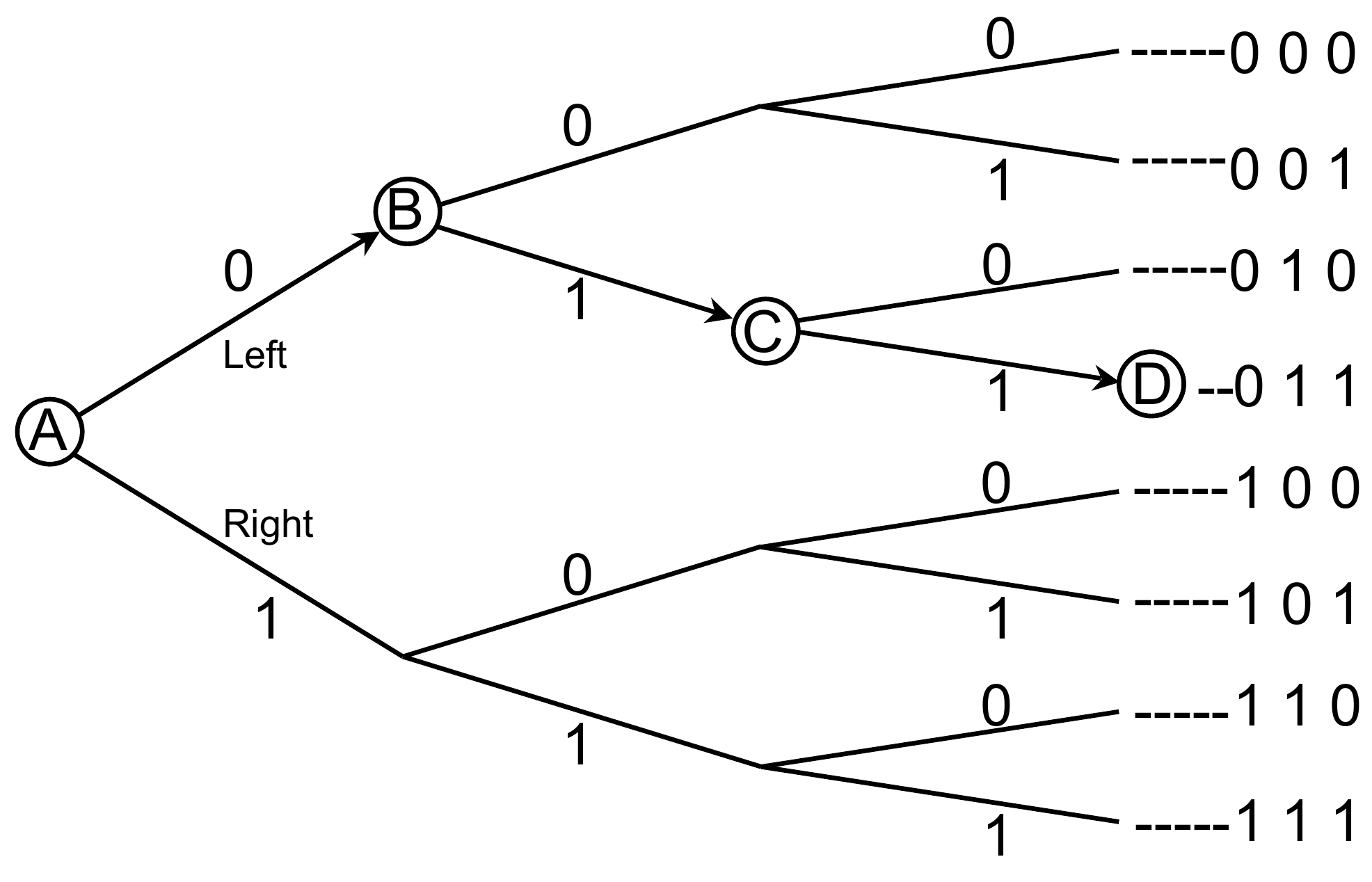} 
\caption{
For a traveller who does not know the way, each fork in the road requires one bit of information to make a correct decision.
The \delete{binary (zero/one) numbers} \edit{0s and 1s} on the right\edit{-}hand side summarise  the instructions needed to arrive at each destination\delete{  if}\edit{;} a left turn is indicated by a 0 and a right turn by a 1.
}
\label{howmanyroads}
\end{center}
\end{figure}
\fi

\section{Finding a Route, Bit by Bit}
\index{bit}%
Information is usually measured in {\em bits}, and one bit of information allows you to choose between two {equally probable}, or {\em equiprobable}, alternatives. 
In order to understand why this is so, imagine you are standing at {the} fork in the road at point A in Figure \ref{howmanyroads}, and that you want to get to the point marked D.  
The fork at A represents two equiprobable alternatives, so if I tell you to go left then you have received one bit of information. If we represent my  instruction with a {\em binary digit} (0=left and 1=right) then this binary digit  provides you with one bit of information, which tells you which road to choose.

Now imagine that you 
come to another fork, at point B in Figure \ref{howmanyroads}. Again, 
a binary digit (1=right) provides one  bit of information,  allowing you to choose the correct road, which leads to C. 
Note that C is one  of {four} possible interim destinations that you could have reached after making two decisions. 
The two binary digits that allow you to make the correct decisions provided two bits {of information}, allow{ing} you to choose from  four (equiprobable) alternatives; 4 equals $2\times2=2^{2}$.

A third binary digit (1=right) provides you with one more bit of information, which allows you to {again} choose the correct road, leading to the point marked D. 
There are now eight roads you could have chosen from when you started at A,
so three  binary digits (which provide you with {three} bits {of information}) allow you to choose from  eight equiprobable alternatives, which also equals $2\times2\times2=2^{3}=8$.

We can  restate this in more general terms if we use $n$ to represent the number of forks, and $m$ to represent the number of final destinations. If you have
come to $n$ forks then you have effectively  chosen from  
$m = 2^{n}$ { final destinations}. 
Because the decision at each fork requires one bit of information, $n$ forks require $n$ bits of information. 

Viewed from another perspective, if there are $m\,=\,8$ possible destinations then the number of forks 
 is $n\,=\,3$, which is the {\em logarithm} of $8$.
\index{logarithm}
Thus, $3\, =\, \log_2 8$ is the number of forks implied by eight destinations. 
More generally, the logarithm of $m$ is the power to which 2 must be raised in order to obtain $m$; that is, $m\,=\,2^n$. Equivalently, given a number $m$, which we wish to express as a logarithm, 
	$n\,=\, \log_2 m. $ 
The subscript {$_2$} indicates that we are using logs to the base 2  (all logarithms in this book {use} base 2  unless stated otherwise).

\section{Bits Are Not Binary Digits} \label{secnotinformation} 
\index{information}%
\index{binary!digits vs bits}%
The word {\em bit} is derived from {\em binary digit}, but a bit and a binary digit are fundamentally different types of quantities. 
A binary digit is the value of a binary variable, whereas a bit is an {\it amount of information}.
To mistake a binary digit for a bit is a category error. In this case, the category error is not as bad as mistaking marzipan for justice, but it is analogous to mistaking a pint-sized bottle for a pint of milk. Just as a bottle can contain between zero and one pint, so a binary digit (when averaged over both of its possible states) can convey between zero and one bit of information. 


\ifFIG
\begin{figure}[b!] 
\begin{center}
\subfloat[]{ \includegraphics[width=0.5\textwidth, angle=0] {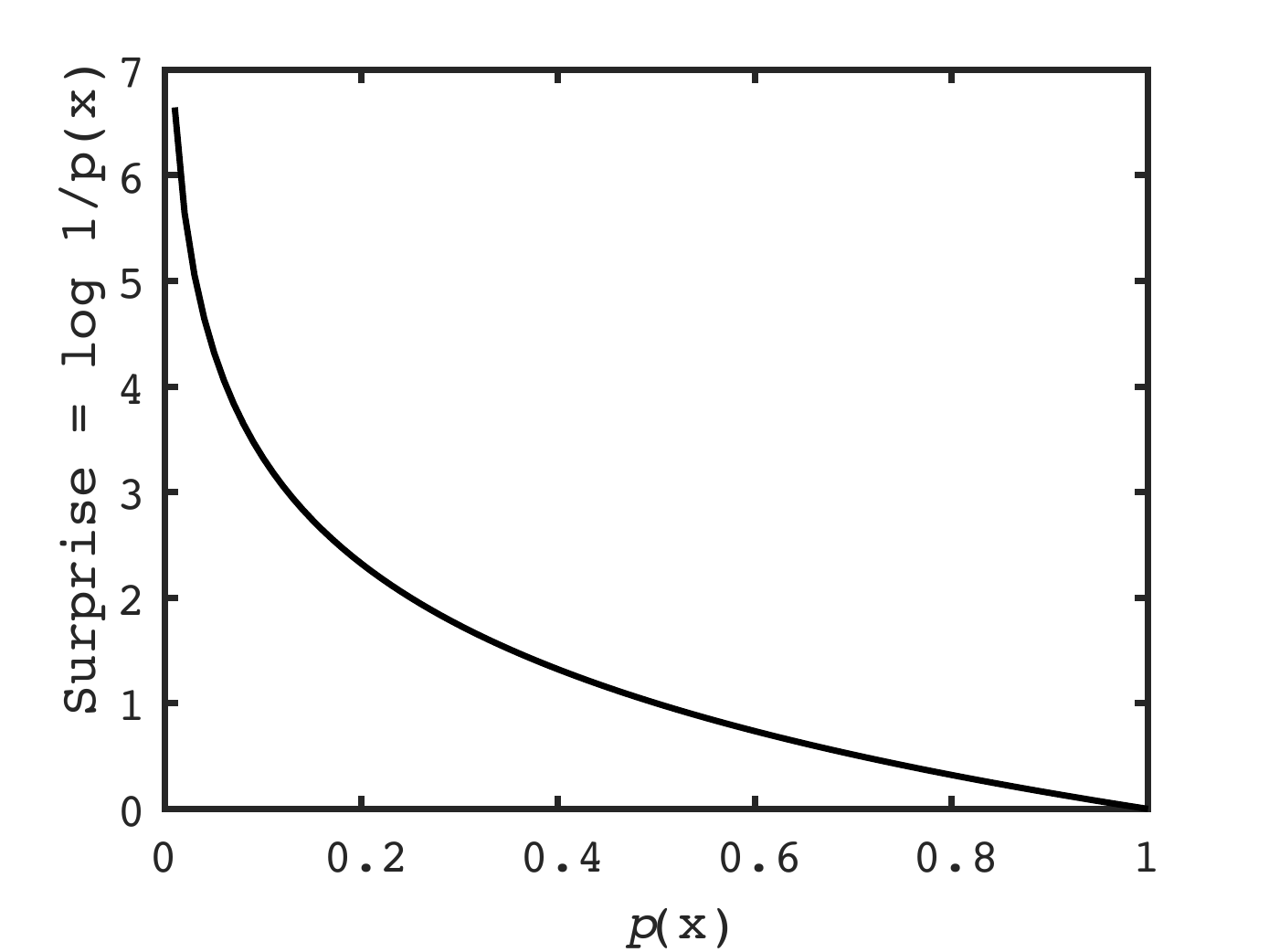} }
\subfloat[]{ \includegraphics[width=0.5\textwidth ] {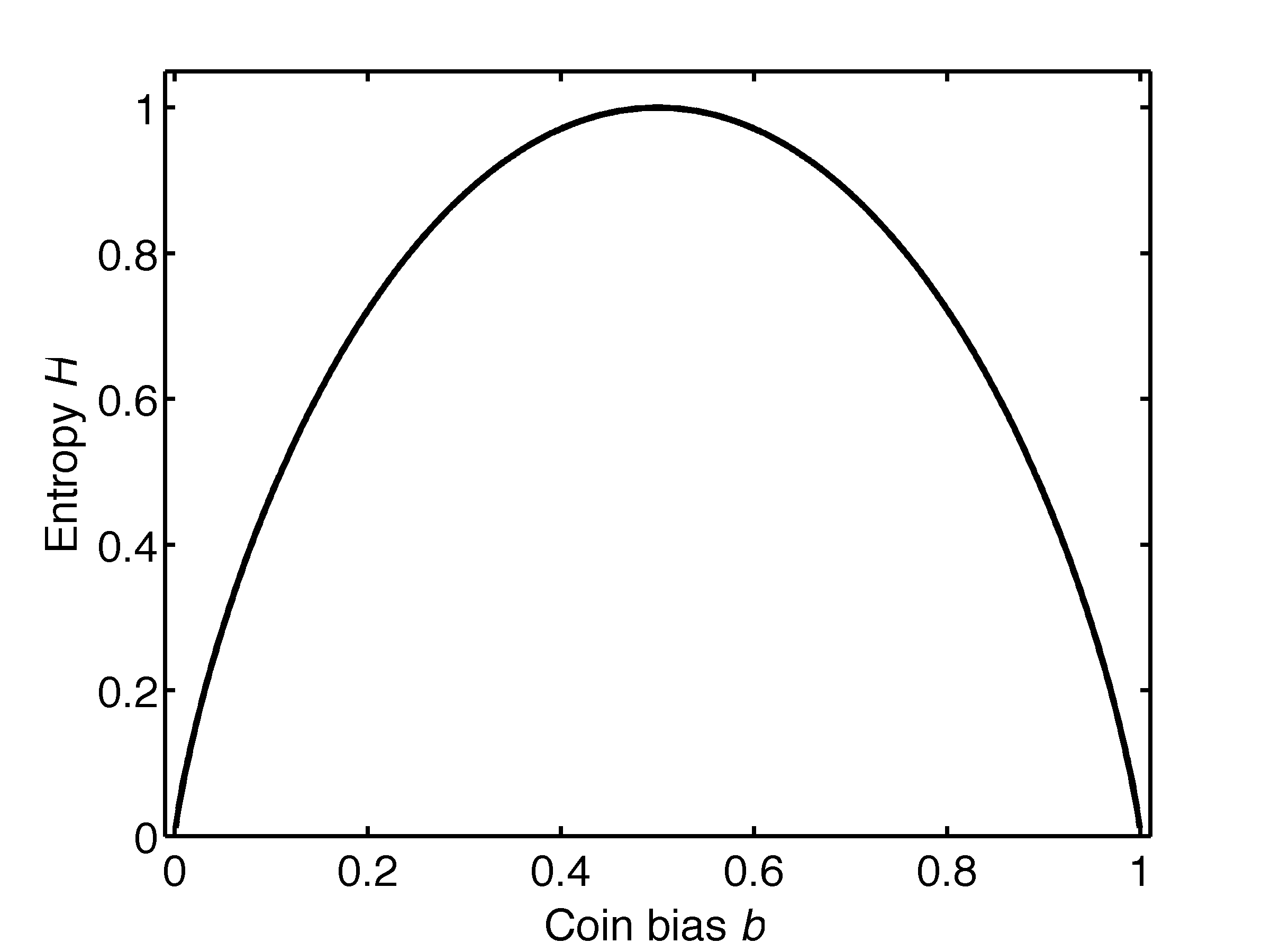} }
\caption{
a) Shannon information as surprise. 
Values of $x$ that are less probable have 
larger values of surprise, defined as $\jlog_{2} (1/p(x))$ bits.
b) Graph of entropy $H(\X)$ versus coin bias (probability $p(\xh)$ of a head).  
The entropy of a coin is the average  amount  of surprise or Shannon information in the distribution of possible outcomes (i.e. heads and tails).
}
\label{figSurprisalBody}\label{figEntropy}
\end{center}
\end{figure}
\fi

\section{Information and Entropy} \label{secise}  \label{sechowsurprised}
Consider a coin which lands heads up 90\% of the time (i.e. $p(x_h)\,=\,0.9$). 
When this coin is flipped, we expect it to land heads up $(x\,=\,x_h)$, \edit{so when it does so} we are less surprised than when it lands tails up ($x\,=\,x_t$).  The more improbable a particular outcome is, the more surprised we are to observe it. 
\index{logarithm}%
\index{bit}%
If we use logarithms to the base 2 then the Shannon information or {\em surprisal} of each outcome  
 is measured in bits (see Figure \ref{figSurprisalBody}a) 
\bea
	{\rm Shannon \: information} & = & \jlog \, \frac{1}{p(x_h)}  \:  \mbox{bits}, \label{eqentropy}
\eea
which is often expressed as: information $=\:-\jlog \: p(x_h) \:  \mbox{bits}$.  

\jsubsection{Entropy is Average Shannon Information}
We can represent the outcome of a coin flip as the {\em random variable} $x$, such that a head is $x\,=\,x_{h}$
and a tail is $x\,=\,x_{t}$.    
In practice, we are not usually interested in the surprise
\index{entropy}
 of a particular value of a {random variable},
 \index{random variable}%
 but we are interested in how much surprise, on average, is associated with the entire set of possible values. \delete{In other words} The average surprise of a variable $\X$ is defined by its probability distribution $p(\X)$, and is called the {\em entropy} of $p(\X)$, represented as $H(\X)$.

\jsubsection{The Entropy of a Fair Coin}
The average amount of surprise 
about the possible outcomes of a coin flip can be found as follows. 
If a coin is fair or unbiased then $p(x_{h})\,=\,p(x_{t})\,=\,0.5$ then the  Shannon information gained when a head or a tail is observed is $ \jlog {1}/{0.5} \,=\,  1 {\rm \: bit}$, so  the average Shannon information gained after each coin flip is also 1 bit.
\index{uncertainty}%
Because entropy is defined as average Shannon information, the entropy of a fair coin is $H(x)=1$ bit.

\ifFIG
\begin{figure}[b] 
\begin{center}
\subfloat[]{\includegraphics[height =0.3 \textwidth] {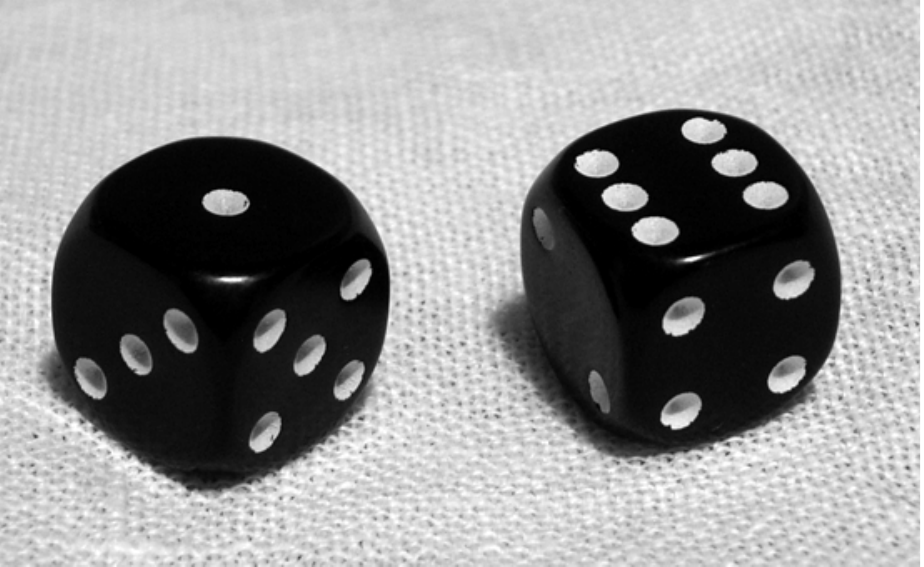}}
\hspace{0.2in}
\subfloat[]{\includegraphics [height =0.31 \textwidth] {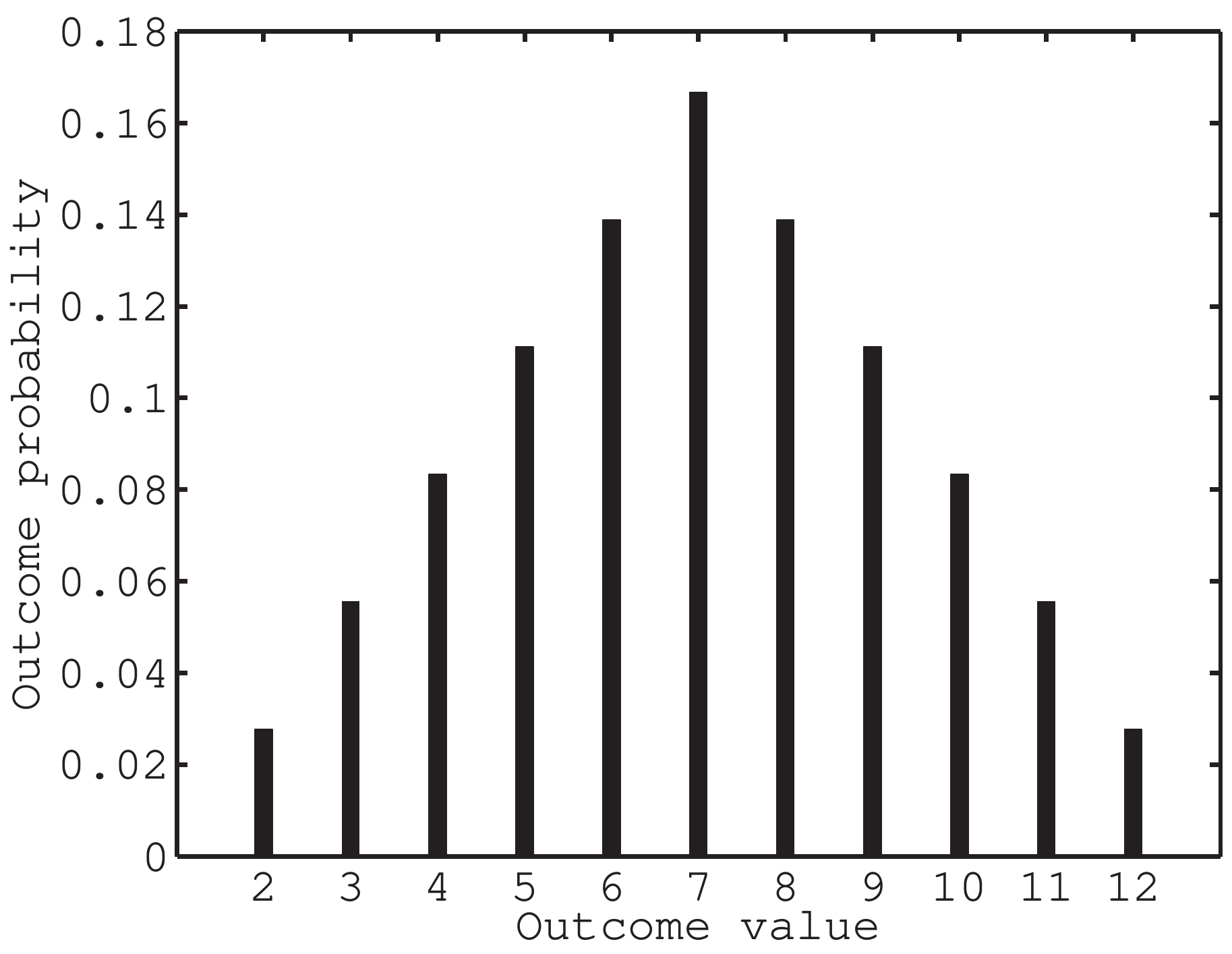} }
\caption{\edit{(}a) A pair of dice. \edit{\quad(}b) Histogram of dice outcome values.  
}
\label{figMacroMicroZ}
\end{center}
\end{figure}
\fi

\jsubsection{The Entropy of an Unfair (Biased) Coin}
If a coin is biased such that 
the probability of a head is $p(x_{h})\,=\,0.9$ then it is easy to predict the result of each coin flip  (i.e. with 90\% accuracy if we  predict a head for each flip). If the outcome is a head then the amount of Shannon information gained  is 
$\jlog (1/0.9) \,=\, 0.15$ bits.
But  if the  outcome is a tail then   the amount of Shannon information gained is
 $ \jlog (1/0.1) \,=\, 3.32$ bits.   
Notice {that} more information is associated with the more surprising outcome. 
Given that the proportion of flips that yield a head is  $p(\xh)$, and that the proportion of flips that yield a tail is  $p(\xt)$
(where $p(\xh)+p(\xt)\,=\,1$), the average  surprise is 
\bea
	H(x) & = &  p(\xh) \log \frac{1}{p(\xh)} + p(\xt) \log \frac{1} {p(\xt)}, \label{eqentrunfaircoin} 
\eea
which comes to $H(x) \,=\, 0.469  \: {\rm bits}$, as in Figure \ref{figEntropy}b. 
If we define a tail as $x_{1}\,=\,\xt$ and a head as $x_{2}\,=\,\xh$ then Equation \ref{eqentrunfaircoin} can be written as
\bea
	 H (\rvs) &  = &  \sum_{i=1}^{2} \: p(x_{i}) \: \jlog \frac{1}{p(x_{i})}  {\rm bits}.   \label{coinentropyabc} \label{eqgenenteropy} 
\eea
More generally, a random variable $\rvs$ with a probability distribution  
\index{probability!distribution}%
\index{entropy}%
$p(\X) \,=\,  \{p(x_1), \dots, p(x_m) \}$ 
has an {entropy} of 
\bea
	 H (\rvs) & = & \sum_{i=1}^{\nc} \: p(\s_{i}) \: \jlog \frac{1}{p(\s_{i})} \bits \label{EntropyDefDiscrete}.
\eea
The reason this definition matters is because Shannon's source coding theorem (see Section \ref{secchannelcap}) guarantees that each value of the variable $\rvs$ can be represented with an average of (just over) $H(\rvs)$ binary digits. However, if the values of consecutive values of a random variable are not independent then each value is more predictable, and therefore less surprising, which reduces the information-carrying capability (i.e. entropy) of the variable.  This is why it is important to specify whether or not consecutive variable values are {\em independent}.
\index{independence}%
\begin{center}
\begin{table*}[b!] 
{
\centering
\begin{tabular} {|c|c |c |c |c |c |}
\hline
Symbol & Sum & 	Outcome 	 	& Frequency & 	$P$ & Surprisal   \\
\hline
$x_1$ & 2 	& 1:1						& 1	& 0.03 & 5.17 	 \\
$x_2$ & 3 	& 1:2, 2:1					& 2	& 0.06 & 4.17 	\\
$x_3$ & 4		& 1:3, 3:1, 2:2				& 3	& 0.08 &  3.59 	 \\
$x_4$ & 5		& 2:3, 3:2, 1:4, 4:1			& 4	& 0.11 & 3.17	 \\
$x_5$ & 6		& 2:4, 4:2, 1:5, 5:1, 3:3		& 5	& 0.14 & 2.85 	 \\
$x_6$ & 7		& 3:4, 4:3, 2:5, 5:2, 1:6, 6:1 	& 6 	& 0.17 & 2.59  \\
$x_7$ & 8		& 3:5, 5:3, 2:6, 6:2, 4:4		& 5	& 0.14 & 2.85 \\
$x_8$ & 9		& 3:6, 6:3, 4:5, 5:4			& 4	& 0.11 & 3.17  \\
$x_9$ & 10	& 4:6, 6:4, 5:5				& 3	& 0.08 & 3.59  \\
$x_{10}$ & 11	& 5:6, 6:5 					& 2	& 0.06 & 4.17  \\
$x_{11}$ & 12	& 6:6						& 1 	& 0.03 & 5.17  \\
\hline
\end{tabular}
}
\caption{A pair of dice have 36 possible outcomes.
\nl
Sum: outcome value, total number of dots for a given throw of the dice. 
\nl
Outcome: ordered pair of dice numbers that could generate each symbol. 
\nl
Freq: number of different outcomes that could generate each outcome value. 
\nl
$P$: the probability that the pair of dice yield a given outcome value (freq/36).
\nl
Surprisal: $P \log (1/P)$ bits.
}
\label{diceA}
\end{table*}
\end{center}
\jsubsection{Interpreting Entropy}
If $H(\X)\,=\,1$ bit  
 then the variable $\X$ could be used to represent 
 $m \,=\,  2^{H(x)} 
 {\rm or \: 2 \: equiprobable \: values}$. Similarly, if $H(\X)\,=\,0.469$ bits 
 then the variable $\X$ could be used to represent 
 $m \,=\,  2^{0.469}$ {or 1.38 equiprobable values}; as if we had a die with 1.38 sides.  
At first sight, this seems like an odd statement. 
Nevertheless, translating entropy into an equivalent number of equiprobable values serves as an  intuitive guide for the amount of information represented by a variable.  

\jsubsection{Dicing With Entropy} \label{secHuffman}
\index{die}%
Throwing a pair of 6-sided dice yields an {\em outcome} in the form of an ordered pair of numbers,
and there are a total of 36 equiprobable outcomes, as shown in Table \ref{diceA}.  If we define an {\em outcome value} as the sum of this pair of numbers then there are $\nc=11$ possible outcome values $A_{x}=  \{2,3,4,5,6,7,8,9,10,11,12 \}$, represented by the symbols $x_1, \dots, x_{11}$.
These outcome values occur with the frequencies shown in Figure \ref{figMacroMicroZ}b and Table \ref{diceA}. Dividing  the frequency  of each  outcome value by 36 yields the probability $P$ of each  outcome value. Using Equation \ref{EntropyDefDiscrete}, we can use these 11  probabilities to find the entropy 
\bea 
	 H(x) &=& p(x_{1}) \log \frac{1}{p(x_{1})} +   p(x_{2})\log \frac{1}{p(x_{2})}  + \dots +  p(x_{11})\log \frac{1}{p(x_{11})}  \nn \\
		&=& 3.27 \bits. \nn
\eea
Using the interpretation described above, a variable with an entropy of 3.27 bits can represent  $2^{3.27} \,=\, 9.65$
equiprobable values. 

\jsubsection{Entropy and Uncertainty}  \label{secmaxent33}
Entropy is a measure of {\em uncertainty}. 
\delete{Crucially,} \edit{W}hen our uncertainty\index{uncertainty} is reduced, we gain {information}\index{information}, so \delete{that} information and entropy are two sides of the same coin.  
However, information has a rather subtle interpretation, which can easily lead to confusion.   
\index{information vs entropy}%
\index{entropy vs information}%

Average information shares the same definition as entropy, but whether we call a given quantity information or entropy depends on whether it is being given to us or taken away. For example, if a variable has high entropy then our initial uncertainty about the value of that variable is large and is, by definition, exactly equal to its entropy. If we are told the value of that variable then, on average, we have been given 
information equal to the uncertainty (entropy) we had about its value. Thus, receiving an amount of information is equivalent to having exactly the same amount of entropy (uncertainty) taken away. 

\section{Entropy of Continuous Variables}  
For discrete variables, entropy is well-defined. However, for all continuous variables, entropy is effectively infinite. 
Consider the difference between a discrete variable $x_{d}$ with $n$ possible values and a continuous variable $x_{c}$ with an uncountably infinite number of possible values; for simplicity, assume that all values are equally probable.
The probability of observing each value of the discrete variable is $P_{d}=1/m$, so the entropy of $x_{d}$ is
$H(x_{d})=\log m$. In contrast, the 
 probability of observing each value of the continuous variable is $P_{c}=1/\infty=0$, so the entropy of $x_{c}$ is
$H(x_{c})=\log \infty = \infty$.  In one respect, this makes sense,  because each value of of a continuous variable is implicitly specified with infinite precision, from which it follows that the amount of information conveyed by each value is infinite. However, this result implies that all  continuous variables have the same entropy. In order to assign different values to different variables,  all  infinite terms are simply ignored,  which yields the {\em differential entropy} 
\bea
	H(x_{c}) & = & \int p(x_{c}) \log \frac{1}{p(x_{c})} \: dx_{c}.
\eea
It is worth noting that the technical difficulties associated with entropy of continuous variables disappear for quantities like mutual information, which involve the difference between two entropies. For convenience, we use the term entropy for both continuous and discrete variables below. 

\section{Maximum Entropy Distributions}  \label{secmaxent}
A distribution of values that has as much entropy (information) as theoretically possible is a {\em maximum entropy distribution}. Maximum entropy distributions are important because, if we wish to use a variable to transmit as much information as possible then we had better make sure it has maximum entropy.  
For a given variable,    the precise form of its maximum entropy distribution depends on the constraints placed on the values of that variable\cite{REZA_INFO_THEORY}. 
It will prove useful to summarise three important maximum entropy distributions.  These are listed in order of decreasing numbers of constraints below.

\jsubsection{The Gaussian Distribution} If a variable $x$ has a fixed variance,  but is otherwise unconstrained, then the maximum entropy distribution is the Gaussian distribution (Figure \ref{fig_exp1}a).  
This  is particularly important in terms of energy efficiency because no other distribution can provide as much information at a lower energy cost per bit. 
\ifFIG 
\begin{figure}[b!] 
\begin{center}
\subfloat[]{\includegraphics [height =0.23 \textwidth] {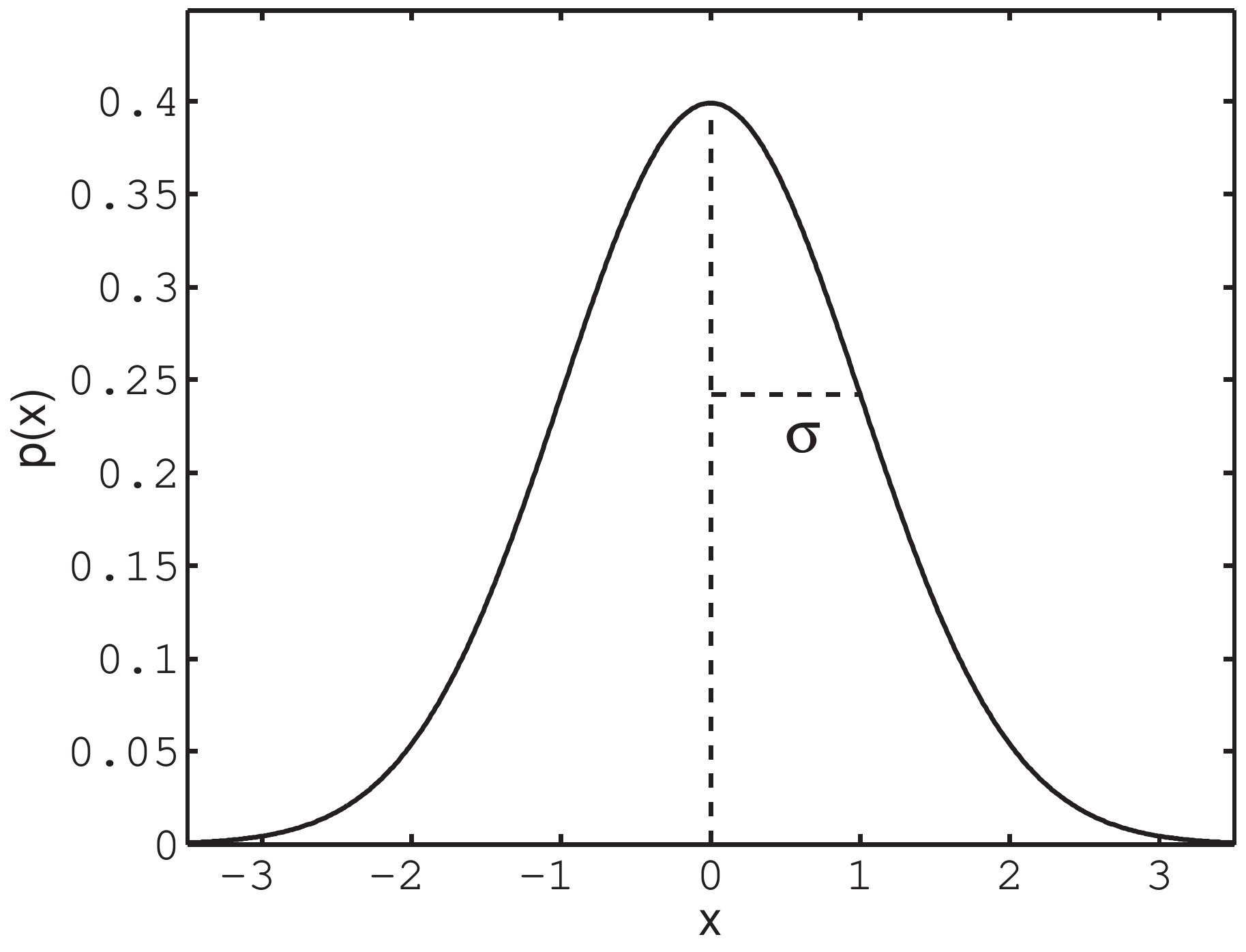} }
\subfloat[]{\includegraphics[height =0.23\textwidth] {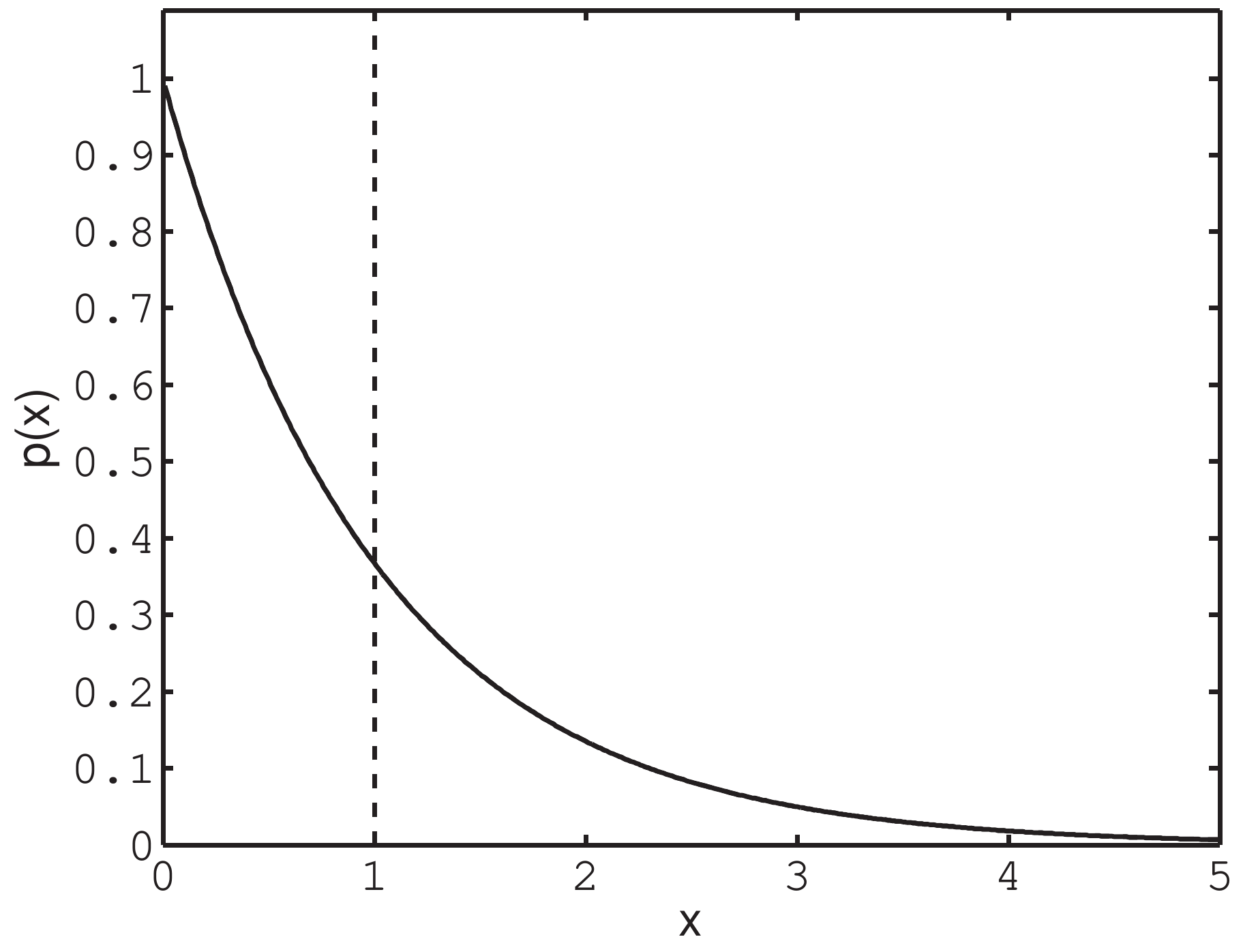}}
\subfloat[]{\includegraphics [height =0.23 \textwidth] {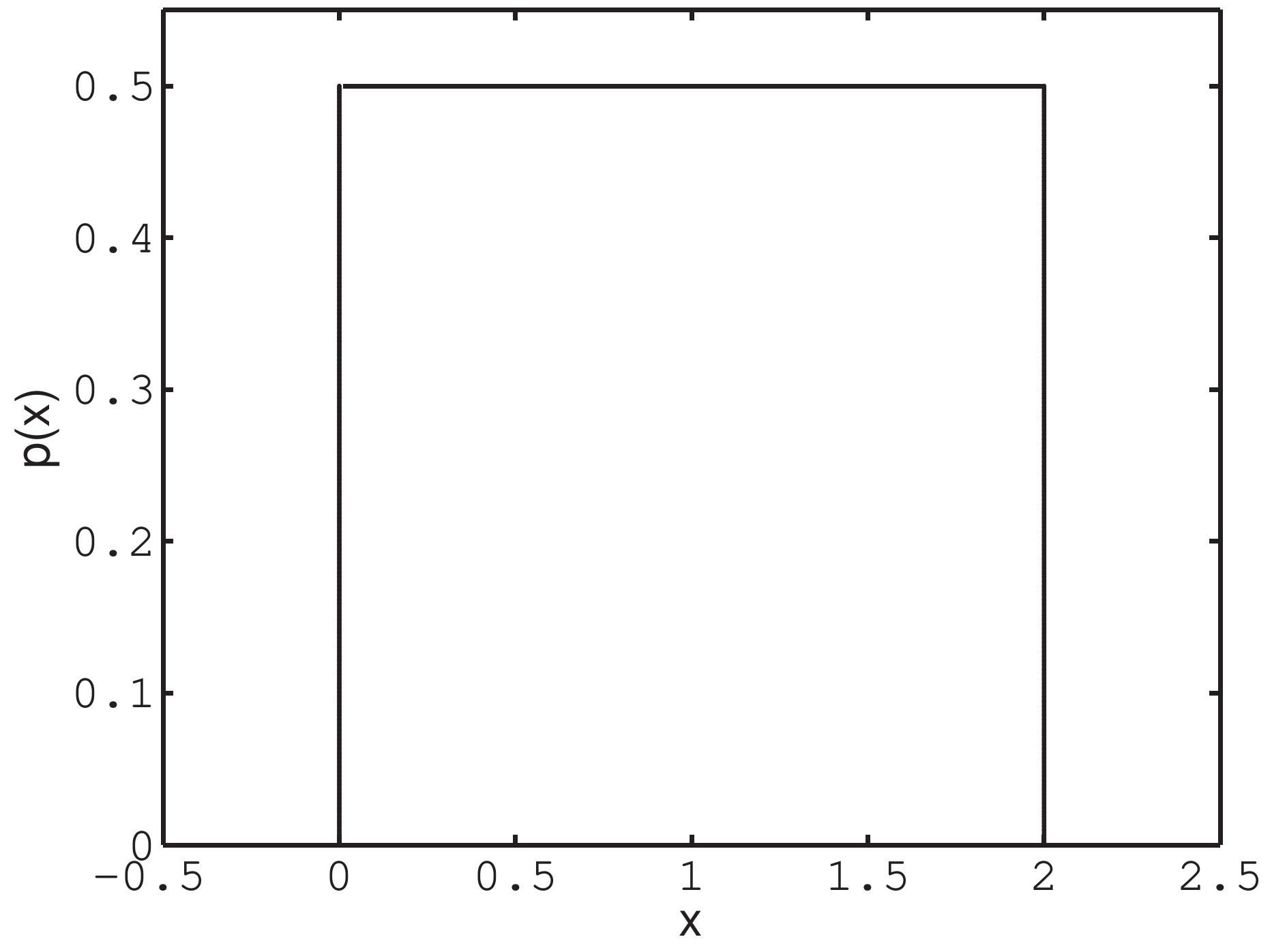} }
\caption{
Maximum entropy distributions. 
a) Gaussian distribution, with mean $\mu\,=\,0$ and a standard deviation  $\sigma\,=\,1$ (Equation \ref{eqgauss}). 
b) Exponential distribution, with mean indicated by the vertical line (Equation \ref{eqexp65}). 
c) 
A uniform distribution with a range between zero and two (Equation \ref{eqexp65AA}). 
}
\label{fig_exp1}
\end{center}
\end{figure}
\fi
If a variable has a Gaussian or {\em normal} distribution then the probability  of observing a particular value $x$ is
\bea
		p(x) 	& = & \frac{1}{ \sqrt{2 \pi v_{x} }} \:
		 e^{- 
		 					(\mu_{x} - x)^{2}  / ({ 2 v_{x} }) 
						},
\label{eqgauss}
\eea 
where $e=2.7183$.   
This equation defines the bell-shaped curve in Figure \ref{fig_exp1}a. The term $\mu_{x}$ is the mean of the variable $x$, and defines the central value of the distribution; we assume that all variables have a mean of zero (unless stated otherwise). The term $v_{x}$ is the variance of  the variable $x$, which is the square of the  standard deviation $\sigma_{x}$ of $x$, and defines the width of the bell curve. Equation \ref{eqgauss} is a {\em probability density function}, and (strictly speaking) $p(x)$ is the {\em probability density} of $x$. 


\index{exponential distribution}%
\jsubsection{The Exponential Distribution}
If a variable has no values below zero, and has a fixed mean $\mu$, 
but is otherwise unconstrained,  then  the maximum entropy distribution is exponential, 
\bea
		p(x)& =& \frac{1}{\mu} e^{- {x}/{\mu}},  \label{eqexp65}
\eea
which has 
 a variance of $\var(x)=\mu^{2}$, as shown in Figure \ref{fig_exp1}b. 

\index{uniform distribution}%
\jsubsection{The Uniform Distribution}
If a variable has a fixed lower bound $x_{min}$ and upper bound $x_{max}$,  but is otherwise unconstrained, then the maximum entropy distribution is uniform,  
\bea
		p(x)& =&{1}/ ({x_{max}-x_{min}}),  \label{eqexp65AA}
\eea
which has a variance $(x_{max}-x_{min})^{2}/12$, as shown in Figure \ref{fig_exp1}c.

\ifFIG
\begin{figure}[b!] %
\begin{center}
\includegraphics[width= 0.4\textwidth] {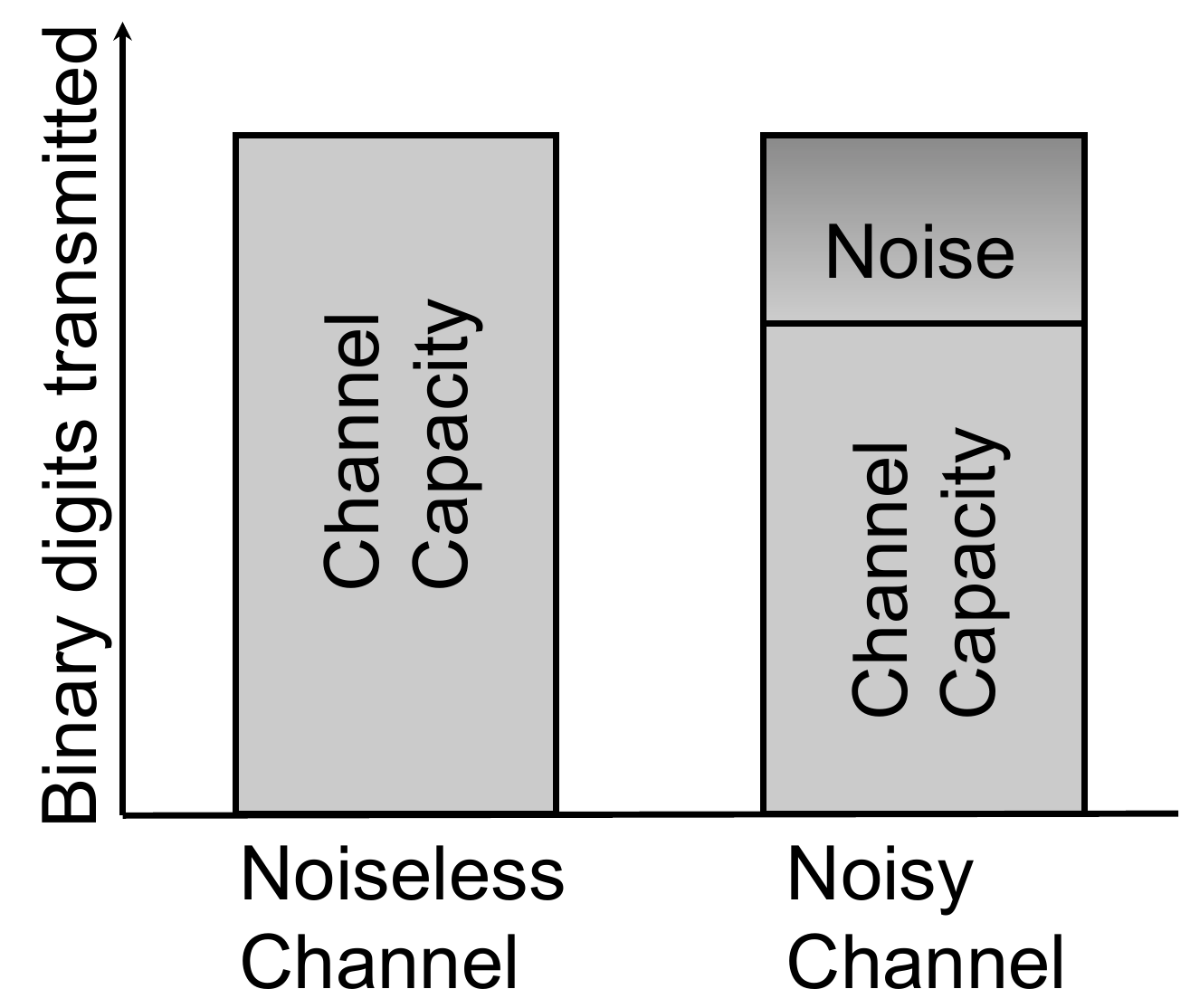} 
\caption{
The {\em channel capacity} of noiseless and noisy channels is the maximum rate at which information can be communicated. 
If a noiseless channel communicates data at 10 binary digits/s then its capacity is $C=10$ bits/s\edit{. T}he capacity of a noiseless channel is numerically equal to the rate at which it communicates binary digits, 
whereas the  capacity of a noisy channel is less than this because it is limited by the noise in the channel. 
}
\label{fignoisefreechannel} 
\end{center}
\end{figure}
\fi

\section{Channel Capacity} \label{secchannelcap}
\index{channel capacity}%
\index{capacity}%

A very important (and convenient) channel is the additive channel. As encoded values pass through an additive channel, noise $\eta$ (eta)  is added, so that the channel output 
is a noisy version $y$ of the channel input $x$
\bea
	y & = & x + \eta.
\eea
\index{channel capacity}%
The {\em channel capacity} $C$ is the maximum amount of information that a channel can provide at its output about the input. 

The rate at which information is transmitted through the channel depends on the entropies of three variables:  
 1) the entropy $H(x)$ of the input, 2)  the entropy $H(y)$ of the output, 
3) the entropy $H(\eta)$ of the noise in the channel.   
If the output entropy is high then this provides a large potential for information transmission, and the extent to which this potential is realised depends on the input entropy and the level of noise. 
{If the noise is low then the output entropy  can be close to the channel capacity}. 
However, channel capacity gets progressively smaller as the noise increases. 
Capacity is usually expressed in bits per usage (i.e. bits per output), or  bits per second (bits/s).

\section{Shannon's Source Coding Theorem} \label{secsourcecodingtheorem}
\index{source coding theorem}%
Shannon's source coding theorem, described below, applies only to noiseless channels. This theorem 
is really about re-packaging (encoding) data before it is transmitted, so that, when it is transmitted, every datum conveys as much information as possible. This theorem is highly relevant to the biological information processing because it defines definite limits to how efficiently sensory data can be re-packaged. {We consider the source coding theorem using binary digits below, but the logic of the argument applies equally well to any channel inputs.}

\index{channel capacity}%
{
Given that a binary digit can convey a maximum of one bit of information, a noiseless channel which communicates $R$  binary digits per second can communicate  information at the rate of up to $R$ bits/s.  
Because the capacity $C$ is the maximum rate at which it can communicate information from input to output, it follows that the capacity of a noiseless channel is numerically equal to  the number $R$ of binary digits communicated per second. However, if each binary digit carries less than one bit (e.g. if consecutive output values are correlated) then the channel communicates information at a lower rate $R \lt C$. 

%
%
%
Now that we \delete{have some} \edit{are} familiar\delete{ity} with the core concepts of information theory, we can quote Shannon's source coding theorem in full. 
This is also known as Shannon's {\em fundamental theorem for a discrete noiseless channel}, 
and as the {\em first fundamental coding theorem}.  
%
\begin{quote}
Let a source have entropy $H$ (bits per symbol) and a channel have a capacity $C$ (bits per second).
Then it is possible to encode the output of the source in such a way as to transmit at the average rate $C/H-\epsilon$ symbols per second over the channel where $\epsilon$ is arbitrarily small. It is not possible to transmit at an average rate greater than $C/H$ [symbols/s]. 
\newline
Shannon and Weaver, 1949\cite{ShannonWeaverBook}. \\ 
\edit{[}Text in square brackets has been added by the author.\edit{]}
\end{quote}
\index{theorem!source coding}
%
%

 %
Recalling the example of the sum of two dice, a naive encoding would require 3.46 ($\log \: 11$) binary digits to represent  the sum of each throw. However, Shannon's source coding theorem guarantees that an encoding exists such that
an average of (just over) 3.27 (i.e. $\log \: 9.65$) binary digits per value of $s$ will suffice (the phrase `just over'  is an informal interpretation of Shannon's more precise phrase `arbitrarily close to').
}

This encoding process yields inputs with a specific distribution $p(\X)$,
{\jvsblack where there are implicit constraints on the form of $p(x)$ (\eg power constraints). 
The shape of the distribution  $p(\X)$ places an upper limit on the entropy $H(\X)$, and therefore on the maximum information that each input can carry.} Thus, the capacity of a noiseless channel is defined in terms of the particular distribution $p(\X)$ which maximises the amount of information per input
\bea
	C  & = &    \max_{\substack{p(\X)}} \: H(\X) \text{ bits per input}. \label{eqtheorem16AAZZ}
\eea
This  
states that channel capacity $C$ is achieved by the distribution $p(\X)$ which makes $H(\X)$ as large as possible
(see Section \ref{secmaxent}).

\section{Noise Reduces Channel Capacity} \label{secMI}
Here, we examine how noise effectively reduces the maximum information that a channel can communicate. 
If the number of equiprobable (signal) input states is $m_{x}$ then the input entropy is 
\bea
	H({x})  &= &\log \: m_{x} \bits.
\eea
For example, suppose there are  $m_{x}\,=\,3$ equiprobable input states, say, $x_{1}\,=\,100$ and $x_{2}\,=\,200$ and $x_{3}\,=\,300$, so the input entropy is
$ 	H({x}) \,=\, \log 3 \,=\, 1.58 \bits$. 
 And if there are $m_{\eta}\,=\,2$ equiprobable values for the channel noise, say, $\eta_{1}\,=\,10$ and $\eta_{2}\,=\,20$, then the noise entropy is $H({\noise})\, =\, \log 2\, =\, 1.00$ bit. 

\ifFIG
\begin{figure}[b!] 
\begin{center}
\includegraphics[width=0.7\textwidth] {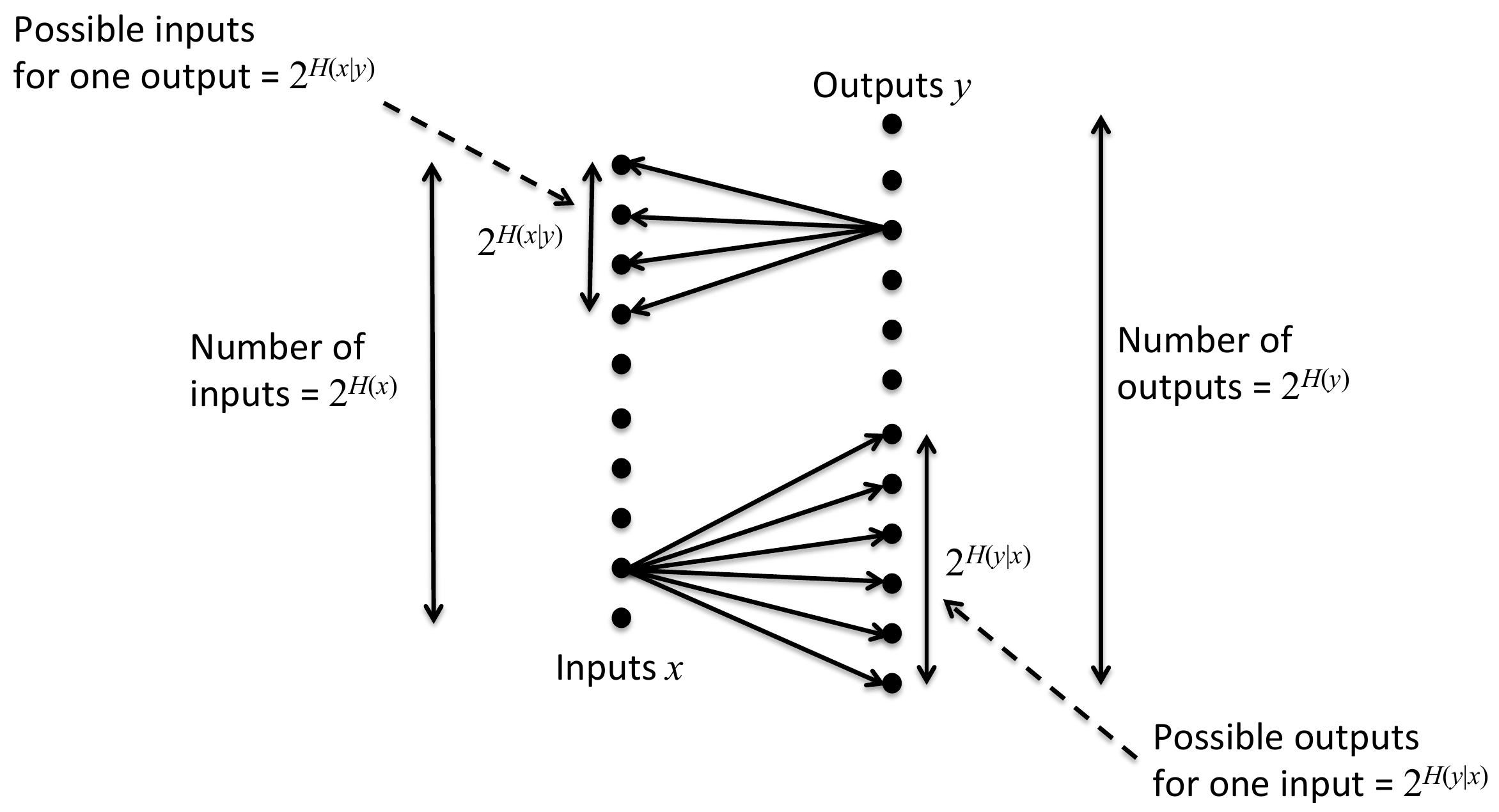} 
\caption{A fan diagram shows how channel noise affects the number of possible outputs given a single input, and {\em vice versa}.
If the  noise $\eta$ in the channel output has entropy  $H(\eta)=H(Y|X)$ then each input value could yield one of $2^{H(Y|X)}$ equally probable output values.
Similarly, if the noise in the channel input has entropy $H(X|Y)$ then each output value could have been caused by one of $2^{H(X|Y)}$ equally probable input values.
}
\label{fignoisecondentropy1}
\end{center}
\end{figure}
\fi

Now, if the input is $x_{1}\,=\,100$ then the output can be one of two equiprobable states, $y_{1}\,=\,100+10=110$ or $y_{2}\,=\,100+20\,=\,120$. And if the input is $x_{2}\,=\,200$ then the output can be either $y_{3}=210$ or $y_{4}=220$. Finally,  if the input is $x_{3}\,=\,300$ then the output can be either $y_{5}\,=\,310$ or $y_{6}\,=\,320$.  
Thus, given three equiprobable input states and two equiprobable noise values, there are $m_{y}\,=\,6(= 3 \times 2)$ equiprobable output states. So the output entropy is 
$H(y) \, = \, \log 6 \, 	= 2.58 \bits$. 
However, some of this entropy is due to noise, so not all of the output entropy comprises {\em information about the input}. 

In general, the total number $m_{y}$ of equiprobable output states is 
$m_{y} \, =\,  m_{x} \times m_{\noise}$, 
from which it follows that the output entropy is 
\bea
	H(y) 
	& = & \log \, m_{x} + \log \, m_{\noise}\\
	& = & H(x) + H(\eta) \bits. \label{eqhycount}
\eea

\ifFIG
\begin{figure}[b!] %
\begin{center}
\includegraphics[width= 0.7\textwidth] {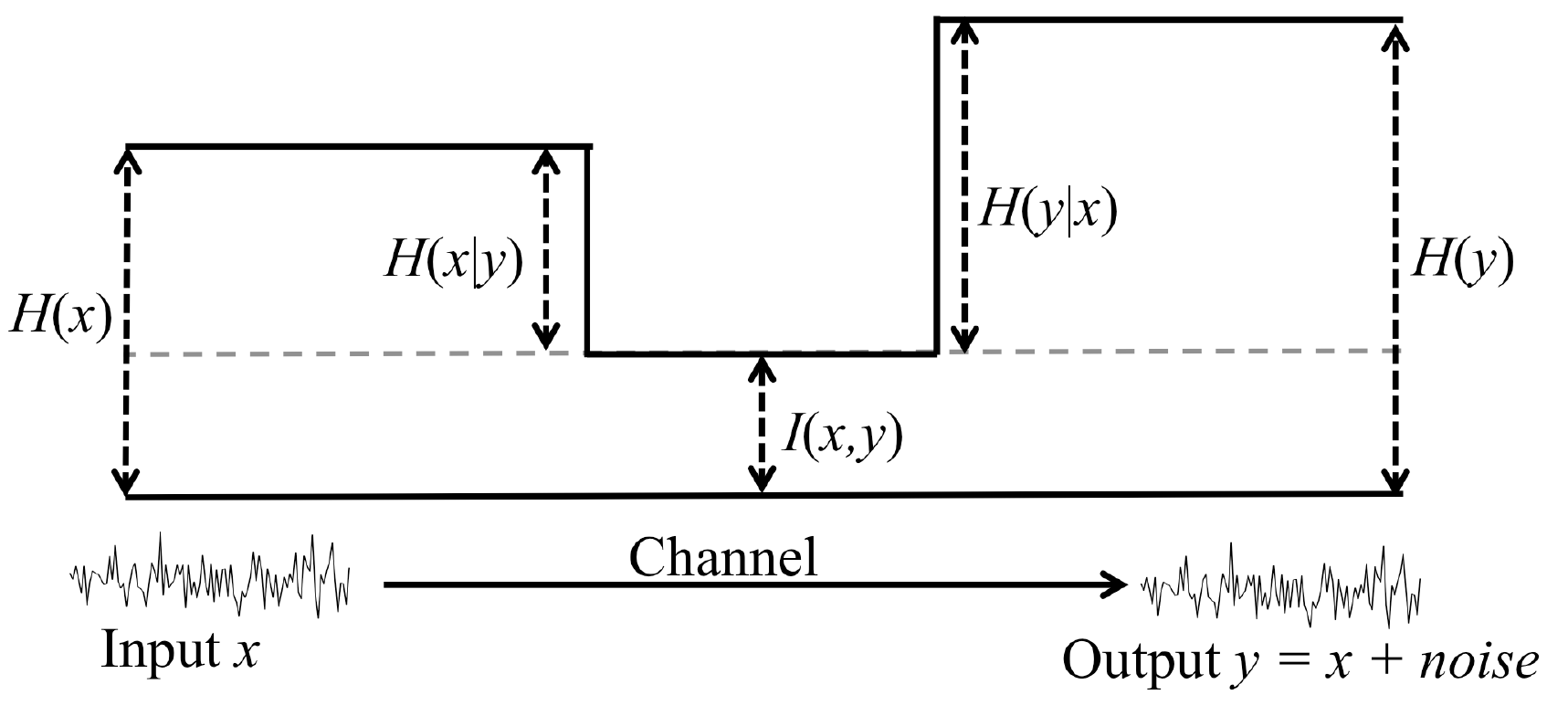} 
\caption{
The relationships between information theoretic quantities. 
Noise refers to noise $\eta$ in the output, which induces uncertainty $H(y|x)=H(\eta)$ regarding the output given the input; this noise also induces uncertainty $H(x|y)$ regarding the input given the output. The mutual information is $I(x,y)\,=\,H(x)-H(x|y)\,=\,H(y)-H(y|x)$ bits.
}
\label{figchannel2017a} 
\end{center}
\end{figure}
\fi

Because we want to explore channel capacity in terms of channel noise, we will pretend to reverse the direction of data along the channel. Accordingly, before we `receive' an input value, we know that the output can be one of 6 values, so our uncertainty about the input value is summarised by its entropy $H({y})\,=\,2.58$ bits. 

\jsubsection{Conditional Entropy} \label{secCE}
{\jvsblack Our average uncertainty about the output value given an input value is the {\em conditional entropy} $H(y|x)$. The vertical bar denotes `given that', so $H(y|x)$ is, `the residual uncertainty (entropy) of $y$ given that we know the value of $x$'.}
\index{conditional entropy}%

After we have received an input value, our uncertainty about the output value is reduced from $H(y)=2.58$ bits to 
\bea
	 H(y|x) \:\:\: = \:\:\:  H(\eta)  \label{eqnoisecond} 
	 \:\:\: = \:\:\: \log 2 \:\:\: = \:\:\: 1 {\rm bit}.
\eea 
Because $H(y|x)$ is the entropy of the channel noise $\eta$, we can write it as $H(\eta)$.  
Equation \ref{eqnoisecond} is true for every input, and it is therefore true for the average input.
Thus, for each input, we gain an average  of 
\bea
	H(y)-H(\eta)  & = & 2.58 - 1 \bits, \label{eqmi3}  \label{eqmi3ab} \label{eqmi3a}
\eea
about the output, which is the {\em mutual information} between $x$ and $y$.
\section{Mutual Information} \label{secMI795}
\index{mutual information}%
The mutual information  $I(x,y)$ between two variables, such as a channel input $x$ and output $y$, is the  average amount of information that each value of $x$ provides about $y$
\bea
	I(x,y) & = &   H(y)-H(\eta)  \bits. \label{eqMI}
\eea
 Somewhat counter-intuitively, the average amount of information gained about the output when an input value is received is the same as the average amount of information gained about the input when an output value is received, $I(x,y) \,=\, I(y,x)$. This is why it did not matter when we pretended to reverse the direction of data through the channel. 
 These quantities are summarised in Figure 
\ref{figchannel2017a}.

\section{Shannon's Noisy Channel Coding Theorem} \label{secsourcecodingtheoremNOISE}
\index{theorem!noisy channel coding}
%
%
All practical communication channels are noisy. To take a trivial example, the voice signal coming out of a telephone is not a perfect copy of the speaker's voice signal, because various electrical components introduce spurious bits of noise into the telephone system. 

As we have seen, the effects of noise can be reduced by using error correcting codes. These codes reduce errors, but they also reduce the rate at which information is communicated. More generally, any method which reduces the effects of noise also reduces \edit{the} rate at which information can be communicated. 
Taking this  line of reasoning to its logical conclusion seems to imply that the only way to communicate information with zero error is to reduce the effective rate of information transmission to zero\edit{, and} \delete{Crucially,} in Shannon's day\delete{,} this was widely believed to be true.
But Shannon proved that information can be communicated, with vanishingly small error, at a rate which is  limited only by the {channel capacity}.

\delete{So, here is}\edit{Now we give} Shannon's {\em fundamental theorem for a discrete channel with noise}, also known as  the {\em second fundamental coding theorem}, and as   {\em Shannon's noisy channel coding theorem}\deletep{, as stated by}  \JVSdelete{Shannon and Weaver (1949)} \deletep{Shannon}\cite{ShannonWeaverBook}:

\begin{quote} 
Let a discrete channel have the capacity $C$ and a discrete source the entropy per second $H$. If $H \leq C$ there exists a coding system such that the output of the source can be transmitted over the channel with an arbitrarily small frequency of errors (or an arbitrarily small  equivocation). If $H \geq C$ it is possible to encode the source so that the equivocation is less than $H-C+\epsilon$ where $\epsilon$ is 
arbitrarily small. There is no method of encoding which gives an equivocation less than $H-C$.
\end{quote} \label{noisycodingtheorem}
%
%
(The word `equivocation'  means the average uncertainty that remains regarding the value of the input after the output is observed\edit{,} \delete{(} i.e.  the conditional entropy $H(X|Y)$\delete{)}). 
In essence,  Shannon's theorem states that it is possible  to use a communication channel
 to communicate information with a low error rate $\epsilon$ (epsilon),  at a rate arbitrarily close to the channel capacity of $C$ bits/s,  
%
but it is not possible to communicate information at a rate greater than  $C$ bits/s.  

%
%

 %
 The capacity of a noisy channel is defined as 
 \bea
	C &  = &    \max_{\substack{p(\X)}} \:\:\: I(x,y) \\
		& = & \max_{\substack{p(x)}} \:\:\: \left[ H(y) - H(y|x) \right]  \text{ bits}.
	\label{eqtheorem16AAZZa}
\eea
If there is no noise (i.e. if $H(y|x)=0$) then  
this reduces to Equation \ref{eqtheorem16AAZZ}, which is the capacity of a noiseless channel. 
The  {\em data processing inequality} states that, no matter how sophisticated any device is, the amount of information $I(x,y)$ in its output about its input cannot be greater than the amount of information $H(x)$ in the input. 
\index{data processing inequality}%

\renewcommand{\X}{{x}}%
\renewcommand{\Y}{{y}}%
\section{The Gaussian Channel} \label{secMIContFixedVariance}
\index{Gaussian channel}%
If the  noise values in a channel are drawn independently from a Gaussian distribution (i.e.  $\eta \sim {\mathcal N}(\mu_{\eta},v_{\eta})$, as defined in Equation \ref{eqgauss}) then this defines a {\em Gaussian channel}.

\ifFIG 
\begin{figure}[b!] 
\begin{center}
{\includegraphics[width =0.5 \textwidth] {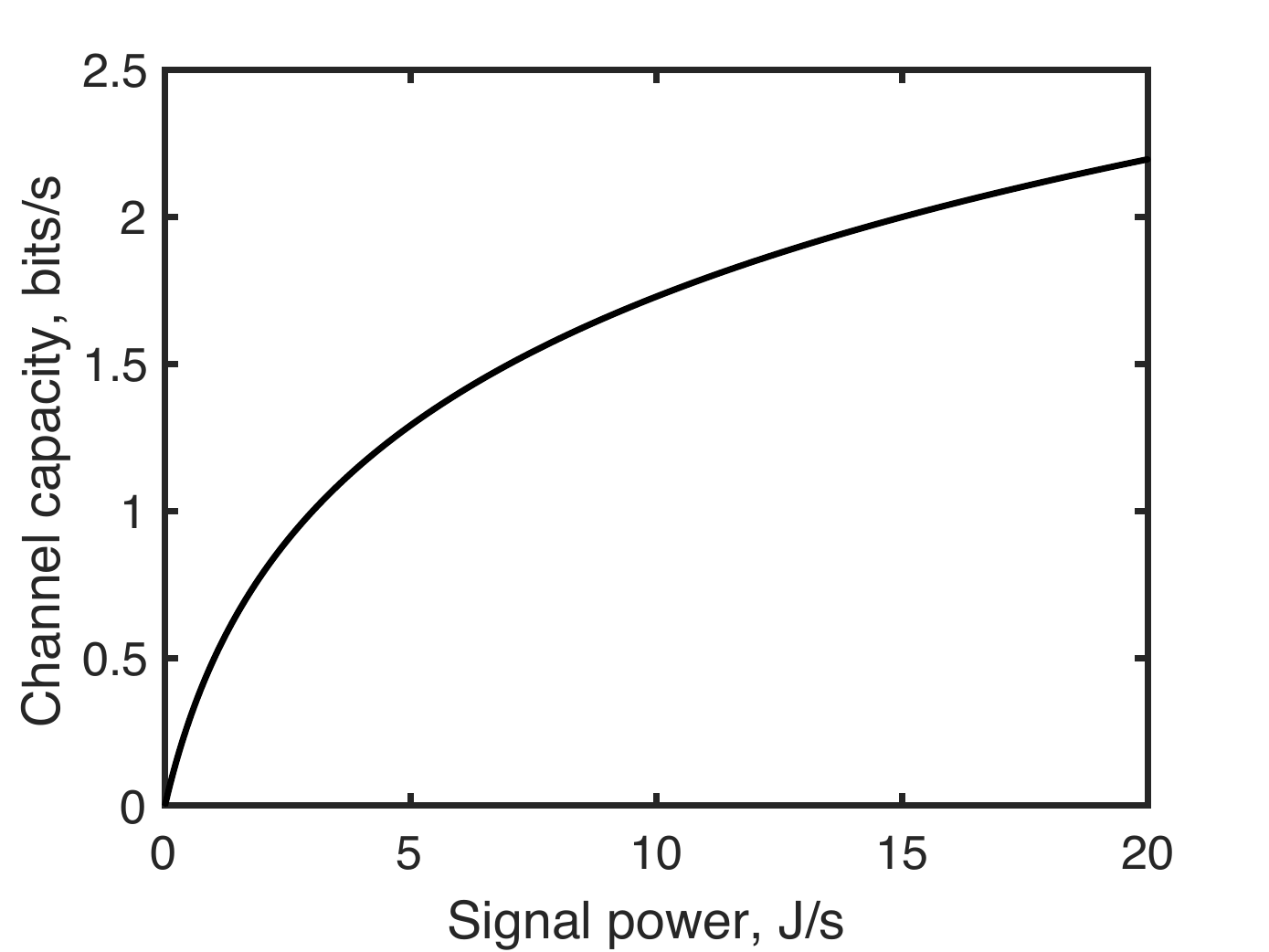}}
\caption{Gaussian channel capacity $C$ (Equation \ref{eqSNRcap}) increases slowly with signal power $S$, which equals signal power here because $N=1$. 
}\label{figsignoisecapacity}
\end{center}
\end{figure}
\fi

Given that $\Y\,=\,\X+\noise$,  if we want $p(\Y)$ to be Gaussian then we should ensure that $p(\X)$ and $p(\noise)$ are Gaussian, because the sum of two independent Gaussian variables  is also Gaussian\cite{REZA_INFO_THEORY}. 
So, $p(\X)$ must be (iid) Gaussian in order to maximise $H(\X)$, which maximises $H(\Y)$, which maximises $I(\X,\Y)$. 
Thus, if each input, output, and noise variable is (iid) Gaussian then the average amount of information communicated per output value is the channel capacity, so that $I(\X,\Y) = C$ bits.
This is an informal statement of  {\em Shannon's  continuous noisy channel coding theorem for Gaussian channels}.  
We can use this to express capacity in terms of the 
 input, output, and noise.  
 
 If the channel input  $x \, \sim \, {\mathcal N}(\mu_{x},v_{x})$  then the continuous analogue (integral) of Equation \ref{EntropyDefDiscrete} yields the {\em differential entropy} 
\index{differential entropy}%
\bea
					H(x) 
	& = & (1/2) \log \: 2\pi e v_x \bits. \label{eqgaussentropyx}
\eea
The distinction between differential entropy effectively disappears when considering the difference between entropies,
and we will therefore find that we can safely ignore this distinction here.  
 Given that the channel noise is iid Gaussian,
 its {entropy} is 
\bea
					H(\noise) & = & (1/2)  \log \: 2\pi e v_\noise \bits. \label{eqgaussentropy1}
\eea
Because the output is the sum  $\Y=\X+\noise$, it is also iid Gaussian with variance  $v_{y} =  v_x+ v_\noise$, and its  entropy is
\bea
					H(\Y) 
					& = &  (1/2) \log \: 2\pi e (v_x + v_{\noise}) \bits.    \label{eqgaussentropyy}\label{eqgaussentropyyA}
\eea
Substituting Equations \ref{eqgaussentropy1} and \ref{eqgaussentropyy} into Equation \ref{eqMI} yields
\bea
	I(x,y) & = & \frac{1}{2} \log   \left( 1+ \frac{v_x}{v_\noise }  \right)  \bits, \label{eqstonnn} \label{eqC236654a} 
\eea
which allows us to choose one out of $m=2^{I}$ equiprobable values. For a Gaussian channel, $I(x,y)$ attains its maximal value of $C$ bits.

The variance of any signal with a mean of zero is equal to its {\em power},  
\index{power}%
which is the rate at which energy is expended per second, and the physical unit of power is measured in {\em Joules} per second (J/s) or {\em Watts}, where 1 Watt = 1 J/s. 
Accordingly, the signal power is $S\, =\, v_x$ J/s, and the noise power is $N \,=\, v_\noise$ J/s. 
This yields   Shannon's famous equation for the capacity of a Gaussian channel
\bea
		C 	& = & \frac{1}{2} \log  \left( 1+\frac{S}{ N} \right)  \text{ bits,} \label{eqSNRcap}
\eea
where the ratio of variances $S/N$ is the {\em signal to noise ratio} (SNR), as in  Figure \ref{figsignoisecapacity}.  
\index{signal to noise ratio}%
It is worth noting that, given a Gaussian signal obscured by Gaussian  noise, the probability of detecting the signal is\cite{SchultzScholarpedia2007} 
\bea
	P & = &  \frac{1}{2} \log \left( 1+   \text{erf}{  \left( \sqrt{\frac{S}{8N}} \right) } \right),
\eea
where erf is the cumulative distribution function of a Gaussian.

\section{Fourier Analysis} \label{secbcpower}
\index{power}%
\index{bandwidth}%
\index{Fourier analysis}%
If a sinusoidal signal has a {\em period} of $\lambda$ seconds then it has a frequency of $f=1/\lambda$ periods per second, measured in {\em Hertz} (Hz).  
A sinusoid with a frequency of $W$\,Hz can be represented perfectly if its value is measured at the {\em Nyquist sample rate}\cite{Nyquist1928}  
of $2W$ times per second. 
\index{Fourier analysis}%
Indeed, {\em Fourier analysis} allows almost any signal $x$ to be represented as a mixture of sinusoidal {\em Fourier components} $x(f): (f=0,\dots,W$),
 shown in Figure \ref{fig_ch0426Fourier3}.  A signal which includes frequencies between 0\,Hz and $W$Hz  has   a {\em bandwidth} of $W$\,Hz. 
\index{bandwidth}%

\ifFIG
\begin{figure}[b!] 
\begin{center}
\includegraphics[width=0.6 \textwidth] {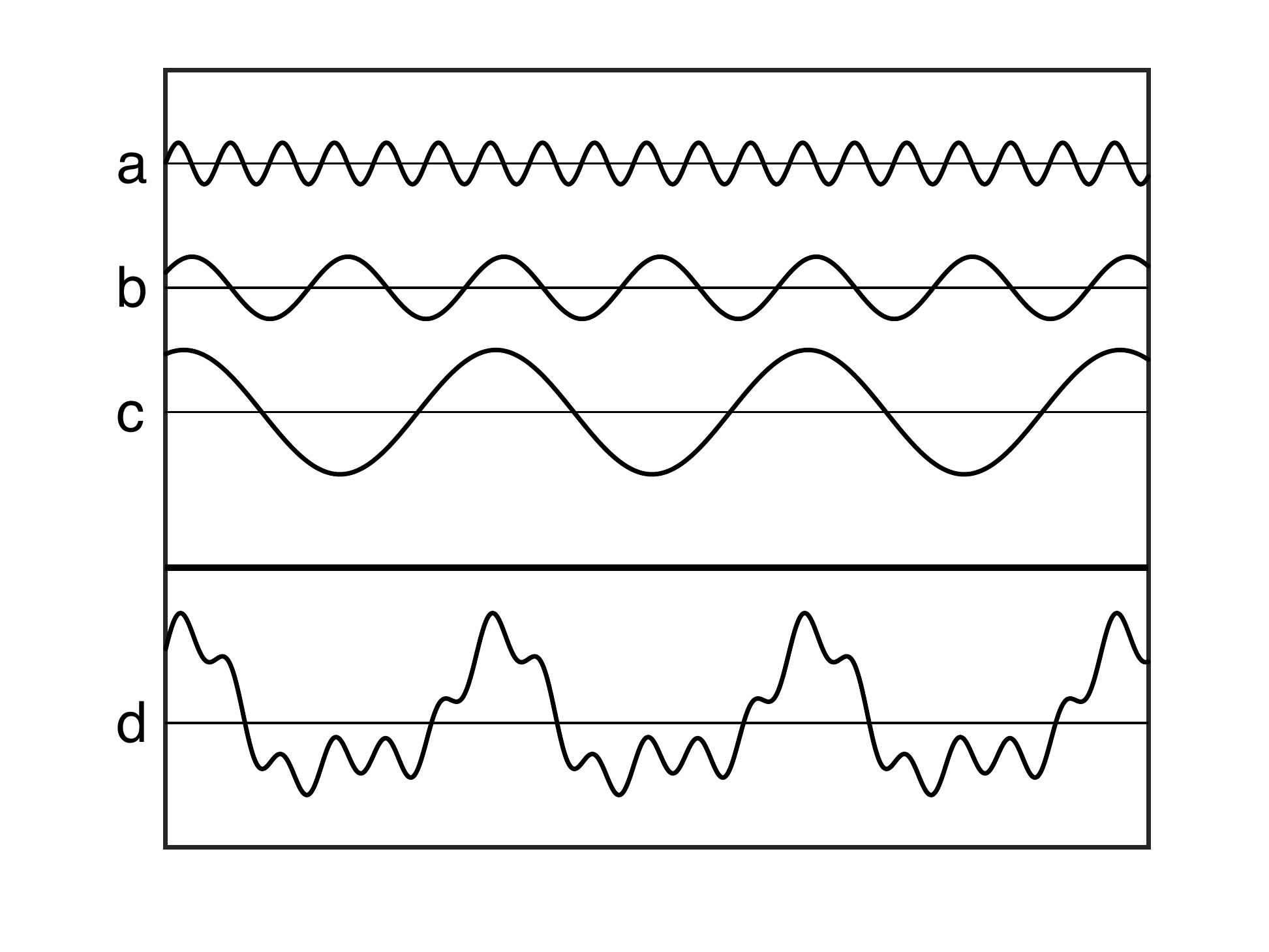} 
\caption{
Fourier analysis decomposes the signal $x$ in d into a unique set of sinusoidal {\em Fourier components} 
$x(f)$ ($f=0,\dots,W$Hz) in a-c, where d=a+b+c. 
}
\label{fig_ch0426Fourier3}
\end{center}
\end{figure}
\fi

\jsubsection{Fourier Analysis} \label{secFFT}
\index{Fourier analysis}%
Fourier analysis allows any signal to be  represented as a weighted sum of sine and cosine functions (see Section \ref{secbcpower}).   
More formally, consider a signal $x$ with a value $x_t$ at time $t$,  which spans a time interval of $T$ seconds. This signal can be represented as a weighted average of sine and cosine functions
\bea
	x_t & = & x_0 \:  + \:  \sum_{n=1}^\infty a_n \cos(f_n t) \: + \: \sum_{n=1}^\infty b_n \sin(f_n t),
\eea
where  $f_n=2\pi n/T$ represents frequency, $a_n$ is the Fourier coefficient (amplitude) of  a cosine with frequency $f_n$, and $b_n$ is the Fourier coefficient of a sine with  frequency $f_n$; and $x_0$ represents the background amplitude (usually assumed to be zero).  Taken over all frequencies, these pairs of coefficients represent the {\em Fourier transform} of  $x$. 
    
The Fourier coefficients can be found from the integrals
\bea
	a_n & = &  \frac{2}{T} \int_0^T x_t \cos(f_n t) \: dt \\  \nn \\ 
	b_n & = & \frac{2}{T} \int_0^T x_t \sin(f_n t) \: dt.
\eea
Each coefficient $a_n$ specifies how much of the signal $x$ consists of a cosine at the frequency $f_n$, and $b_n$  specifies how much consists of a sine. Each pair of coefficients specifies the power and {\em phase}
\index{phase} of one frequency component; the power at frequency $f_{n}$ is $S_{f} =({a_n^{2} + b_n^{2}})$, and the phase is $\arctan(b_{n}/a_{n})$. 
If $x$ has a bandwidth of $W$Hz then its {\em power spectrum} is the set of $W$ values $S_{0},\dots,S_{W}$. 
\index{power spectrum}%

An extremely useful property of Fourier analysis is that, when applied to {\em any} variable, the resultant Fourier components are mutually uncorrelated\cite{BialekBook1996},  
and, when applied to any Gaussian variable, these Fourier components are also mutually independent. 
This means that the entropy of any Gaussian variable can be estimated by adding up the entropies of its Fourier components, 
which can be used to  estimate the mutual information between Gaussian variables.

Consider a variable $y\,=\,x+\eta$, which  is the sum of a Gaussian signal $x$ with variance $S$, and Gaussian noise with variance $N$.
 If the highest frequency in $y$ is $W$\,Hz,  and if values of $x$ are transmitted at the Nyquist rate of $2W$\,Hz,  
then the channel capacity  is $2WC$ bits per second, (where $C$ is defined in Equation \ref{eqSNRcap}). Thus, when expressed in terms of bits per second, this yields a channel capacity of 
\index{Fourier analysis}%
\bea
	C 
	& = &  
	 {W} \log \left( 1+\frac{S}{N} \right) \bitss. 
	 \label{eqstandardSNb}  \label{eqSNRcap3}
\eea
\index{bandwidth}%
If the signal power of Fourier component $x(f)$ is $S(f)$, and the noise power of component $\eta(f)$ is $N(f)$ then the signal to noise ratio is $S(f)/N(f)$ (see Section \ref{secFFT}). 
 The  mutual information at frequency $f$ is therefore
 \bea
 	I(x(f),y(f)) & = &  \log \left(  1 + \frac{S(f)}  {N(f)} \right) \bitss. \label{eqIxydecomp685tBZ}
 \eea
Because the Fourier components of any Gaussian variable are mutually independent, the  mutual information
between Gaussian variables can be obtained by summing $I(x(f),y(f))$ over frequency
\bea
		I(x,y) &  = & \int_{f=0}^{W} I(x(f),y(f)) \: df \:\:\: \bitss. \label{eqIrotateABH}
\eea 
If each Gaussian variable $x$, $y$ and $\eta$ is also iid then
 $I(x,y)\,=\,C$\,bits/s, otherwise $I(x,y) \, \lt \, C$\,bits/s\cite{ShannonWeaverBook}. 
If the peak power at all frequencies is a constant $k$ then it can be shown that  $I(x,y)$ is maximised when $S(f)+N(f)\,=\,k$, which defines a flat power spectrum. 
Finally, if the signal spectrum is sculpted 
so that the signal plus noise spectrum is flat then the logarithmic relation in Equation \ref{eqSNRcap} yields improved, albeit still diminishing, returns\cite{BialekBook1996} $C \propto (S/N)^{1/3}$ bits/s. 

\section{A Very Short History of Information Theory} 
Even the most gifted scientist cannot command an original theory out of thin air. Just as Einstein could not have devised his theories of relativity if he had no knowledge of Newton's work, so Shannon could not have created information theory if he had no knowledge of the work of Boltzmann\edit{~}(1875) and Gibbs\edit{~}(1902) on thermodynamic entropy,   Wiener\edit{~}(1927) on signal processing, Nyquist\edit{~}(1928)
on sampling theory, or Hartley\edit{~}(1928)
on information transmission\cite{PierceInfoTheoryBook1961}.  

Even though Shannon  was not alone in trying to solve one of the key scientific problems of his time 
(i.e.  how to define and measure information), 
he was alone in being able to produce a complete mathematical theory of information\edit{:} a theory that might otherwise have taken decades to construct. In effect, Shannon single-handedly accelerated the rate of scientific progress, and it is entirely possible that, without his contribution, we would still be treating information as if it were some ill-defined vital fluid.

\section{\vspace{-0.1in}Key Equations} \label{AppKeyeqs}
\renewcommand{\Y}{y}

\ni
Logarithms use base 2 unless stated otherwise.

\ni
{\bf {Entropy}}
\index{entropy}%
\bea
	 H (\rvs) & = & \sum_{i=1}^{\nc} \: p(x_{i}) \: \jlog \frac{1}{p(x_{i})} \: \text{ bits} \\
	 H(\rvs) & = & \int_{x} \: p(x) \: \jlog {\frac{1}{p(x)}} \: dx \: \text{ bits}
\eea

\ni
{\bf Joint entropy} 
\bea
	H(\X,\Y) & = & \sum_{i=1}^{\nc} \sum_{j=1}^{\nc} \: p(x_{i},y_{j}) \: \log  \frac{1}{p(x_{i},y_{j})} \: \text{ bits} \\ 
        H (\X,\Y) & = & \int_{\s} \int_{\r} p(\r,\s) \: \jlog \frac{1}{p(\r,\s)} \: d\r \: d\s \: \text{ bits} \\[0.1in]
	H(\X,\Y)  & = & I(\X,\Y) + H(\X|\Y) + H(\Y|\X) \: \text{ bits}  
 \eea

\ni
{\bf Conditional Entropy}
\bea
		H(\rvr|\rvs) & = & 
		 \sum_{i=1}^{\nc} \sum_{j=1}^{\nc} \:  p(x_{i},y_{j}) \: \jlog \frac{1}{p(x_{i}|y_{j})} \: \text{ bits} \\
		H(y|x) & = & 
				 \sum_{i=1}^{\nc} \sum_{j=1}^{\nc} \: p(x_{i},y_{j}) \: \jlog \frac{1}{p(y_{j}|x_{i})} \: \text{ bits} \\
\nn \\
		H(\X|\Y) & = & \int_y \int_x p(x,y) \: \jlog \frac{1}{p(x|y)} \: dx \: dy  \: \text{ bits} \\
\nn \\
	H(\Y|\X) & = &  \int_y \int_x p(x,y) \: \jlog \frac{1}{p(y|x)} \: dx \: dy   \: \text{ bits}
\eea
\bea
	H(\X|\Y) & = & H(\X,\Y) - H(\Y) \: \text{ bits} \\ 
	H(\Y|\X) & = & H(\X,\Y) - H(\X) \: \text{ bits}
\eea
from which we obtain the {\em chain rule for entropy} 
\index{chain rule for entropy}%
\bea
	H(\X,\Y) & = & H(\X) + H(\Y|\X) \: \text{ bits}  \\
	 & = & H(\Y) + H(\X|\Y) \: \text{ bits}
\eea

\ni
{\bf Marginalisation}
\bea
	p(x_{i}) \:=\:  \sum_{j=1}^{\nc} \: p(x_{i},y_{j}), &&
	p(y_{j}) \:= \: \sum_{i=1}^{\nc} \: p(x_{i},y_{j})  \\
	p (x)  \: = \: \int_y p(x,y) \: dy,  &&
	p (y)  \: =  \: \int_x p(x,y) \: dx
\eea
\index{marginalisation}%

\index{mutual information}
\ni
{\bf Mutual Information}\\
\bea
	I(x,y) & = &  
	 \sum_{i=1}^{\nc} \sum_{j=1}^{\nc} \:  p(x_{i},y_{j}) \: \jlog \frac{p(x_{i},y_{j})}{p(x_{i})p(y_{j})}  \: \text{ bits}  \\[0.1in]
	I(\X,\Y) & = & \int_{y} \int_{x} p(x,y) \: \jlog \frac{p(x,y)}{p(x)p(y)} \: dx \: dy \: \text{ bits} \\[0.15in]
	I(\X,\Y) & = & H(x) + H(y) - H(x,y) \label{F18}   \\
	 & = & H(x)-H(x|y)  \\
	 & = & H(\Y)-H(\Y|\X)   \\
	 & = & H(\X,\Y) - [H(\X|\Y) + H(\Y|\X)] \: \text{ bits}
 \eea
If $y\,=\,x+\eta$, with $x$ and $y$ (not necessarily iid) Gaussian variables then
\bea
	I(x,y) & = & 
	\int_{f=0}^{W} \log \left(  1 + \frac{S(f)} {N(f)} \right) \: df  \:\:\:  \text{ bits/s}, \label{eqIxydecomp685tB}
\eea 
\index{Fourier analysis}%
where $W$ is the bandwidth, $S(f)/N(f)$ is the signal to noise ratio of the signal and noise Fourier  components at frequency $f$ (Section \ref{secFFT}), and data are transmitted at the Nyquist rate of $2W$ samples/s. 

\ni
{\bf Channel Capacity}
\bea
	C & = &  \max_{\substack{p(\X)}} \: I(\X,\Y) \text{ bits per value}.  \label{eqtheorem16RPT21}
\eea
\index{capacity}%
If the channel input $x$ has variance $S$, the noise $\eta$ has variance $N$, and both $x$ and $\eta$ are iid Gaussian variables then $I(x,y)=C$,  where 
\bea
	C 	& = &  \frac{1}{2} \log  \left( 1+\frac{S}{N} \right)  \text{ bits per value}, 
\eea
where the ratio of variances $S/N$ is the {signal to noise ratio}.

\section*{Further Reading}

\noindent
Applebaum D (2008)\cite{ApplebaumInfoBook2008}. Probability and Information: An Integrated Approach. 
{\it  A thorough introduction to information theory, which strikes a good balance between intuitive and technical explanations. 
}

\noindent
Avery J (2012)\cite{AveryInformationEvolution2003}. Information Theory and Evolution.
{\it An engaging account of how information theory is relevant to a wide range of natural and man-made systems, including evolution, physics, culture and genetics. Includes interesting background stories on the development of ideas within these different disciplines.
}

\noindent
Baeyer HV (2005)\cite{Baeyer2005}. 
Information: The New Language of Science
{\it  Erudite, wide-ranging, and insightful account of information theory. Contains no equations, which makes it very readable. 
}

\noindent
Cover T and Thomas J (1991)\cite{COVER_THOMAS_INFO}.  Elements of Information Theory.
{\it  Comprehensive, and highly technical, with historical notes and an equation summary at the end of each chapter.
}

 \noindent
Ghahramani Z (2002). Information Theory.
Encyclopedia of Cognitive Science. 
{\it An excellent, brief overview of information.}

 \noindent 
Gleick J  (2012)\cite{GleickInformation}.
The Information. 
{\it  An informal introduction to the history of ideas and people associated with information theory. }

\noindent
Guizzo EM (2003)\cite{GuizzoShannonBio}. The Essential Message: Claude Shannon and the Making of Information Theory.
Master's Thesis, 
Massachusetts Institute of Technology. 
{\it  One of the few accounts of Shannon's role in the development of  information theory.
See} \url{http://dspace.mit.edu/bitstream/handle/1721.1/39429/54526133.pdf}. 

\noindent
Laughlin, SB (2006). The Hungry Eye: Energy, \mbox{Information} and \mbox{Retinal Function},
{\it Excellent lecture on the energy cost of Shannon information in eyes.
See} \url{http://www.crsltd.com/guest-talks/crs-guest-lecturers/simon-laughlin}.

 \noindent 
MacKay DJC (2003)\cite{MacKayInfoTheoryBook}.
Information Theory, Inference, and Learning Algorithms.
{\it  The modern classic on information theory. A very readable text that roams far and wide over many topics.
The  book's web site (below) also has a link to an excellent series of video lectures by MacKay.
Available free online at 
}
\url{http://www.inference.phy.cam.ac.uk/mackay/itila/}. 


%
%

%


\noindent
Pierce JR (1980)\cite{PierceInfoTheoryBook1961}. An Introduction to Information Theory: Symbols, Signals and Noise. Second Edition.
{\it Pierce writes with an informal, tutorial style of writing, but does not flinch from presenting the fundamental theorems of information theory. 
This book provides  a good balance between words and equations. 
}

\noindent
Reza FM (1961)\cite{REZA_INFO_THEORY}. An Introduction to Information Theory. 
{\it A more comprehensive and mathematically rigorous book than  Pierce's book, it should be read only after first reading Pierce's more informal text.
}

\noindent
Seife C (2007)\cite{Seifedecodinguniverse}. 
Decoding the Universe: How the New Science of Information Is Explaining Everything in the Cosmos, From Our Brains to Black Holes.
{\it A lucid and engaging account of the relationship between information, thermodynamic entropy and quantum computing. Highly recommended. 
}

\noindent
Shannon CE and Weaver W (1949)\cite{ShannonWeaverBook}.
The Mathematical Theory of Communication. University of Illinois Press.
{\it 
A surprisingly accessible book,  written in an era when information theory was known only to a privileged few. 
This book can be downloaded from}
 \url{http://cm.bell-labs.com/cm/ms/what/shannonday/paper.html}

\noindent
Soni, J and Goodman, R (2017)\cite{soni2017mind}. 
A mind at play: How Claude Shannon invented the information age
{\it A biography of Shannon. 
}

\noindent
Stone, JV (2015)\cite{StoneInformationBook2014}
Information Theory: A Tutorial Introduction. 
{\it A more extensive introduction than the current article.}
%

For the complete novice, the videos at the online Kahn Academy provide an excellent introduction.
Additionally, the online Scholarpedia web page by Latham and Rudi provides a lucid technical account of mutual information: 
\\ 
\url{http://www.scholarpedia.org/article/Mutual_information}.

Finally, some historical perspective is provided in a long interview with Shannon conducted in 1982:
\url{http://www.ieeeghn.org/wiki/index.php/Oral-History:Claude_E._Shannon}.

 
 \singlespace

\ifFIG
\begin{figure}[b!] 
\begin{center}
\includegraphics[width=0.4 \textwidth] {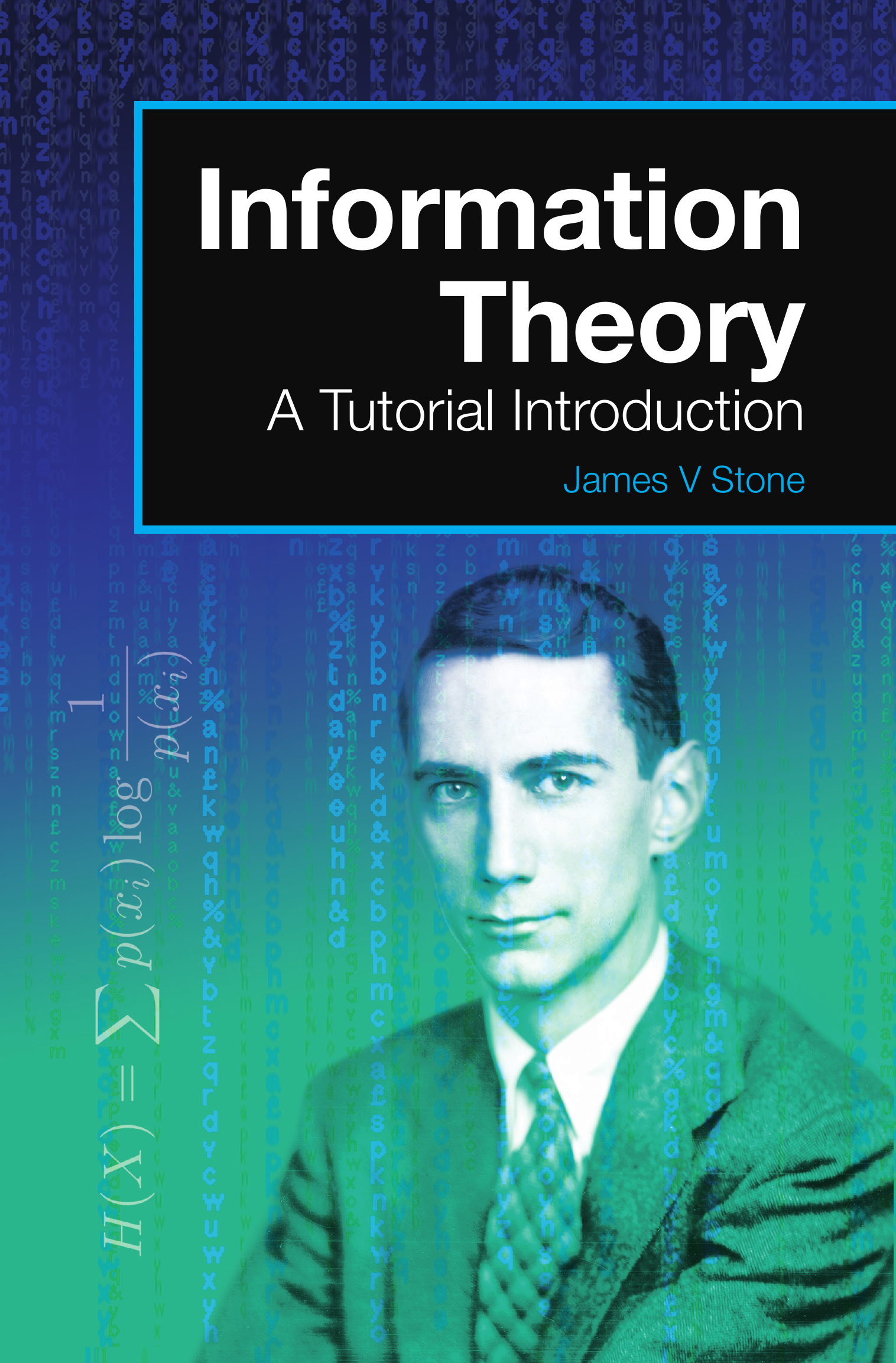} 
\caption{This paper is derived from the book Information Theory. For details, see https://jim-stone.staff.shef.ac.uk/BookInfoTheory/InfoTheoryBookMain.html}
\label{fig_ch0426Fourier3xxx}
\end{center}
\end{figure}
\fi

\end{document}